\documentclass[]{jfm}

\usepackage{graphicx}
\usepackage{newtxtext}
\usepackage{newtxmath}
\usepackage{natbib}
\usepackage{hyperref}
\hypersetup{
    colorlinks = true,
    urlcolor   = blue,
    citecolor  = black,
}

\newcommand{\RomanNumeralCaps}[1]
\author{J. Heyman, E. Villermaux, P. Davy \& T. Le Borgne}
\begin{document}

%\linenumbers
%\preprint{APS/123-QED}

\title{Mixing as a correlated aggregation process}

%\author{J. Heyman$^1$, T. Le Borgne$^1$, P. Davy$^1$, E. Villermaux$^2$}
%\affiliation{$^1$~Geosciences Rennes, UMR 6118, Universit\'{e} de Rennes 1, CNRS, 35042 Rennes, France, \\
%~$^2$~IUSTI, Aix Marseille }
%
%\author{J. Heyman}
%\affiliation{Geosciences Rennes, UMR 6118, Universit\'{e} de Rennes 1, CNRS, 35042 Rennes, France}
%\author{T. Le Borgne}
%\affiliation{Geosciences Rennes, UMR 6118, Universit\'{e} de Rennes 1, CNRS, 35042 Rennes, France}
%\email{joris.heyman@univ-rennes1.fr}
%\author{P. Davy}
%\affiliation{Geosciences Rennes, UMR 6118, Universit\'{e} de Rennes 1, CNRS, 35042 Rennes, France}
%\author{E. Villermaux}
%\affiliation{IUSTI, Aix Marseille }

%\pacs{}

\maketitle

%\tableofcontents

\begin{abstract}
Mixing describes the process by which solutes evolve from an initial heterogeneous state to uniformity under the stirring action of a fluid flow. Fluid stretching forms thin scalar lamellae which coalesce due to molecular diffusion. Owing to the linearity of the advection-diffusion equation, coalescence can be envisioned as an aggregation process. Here, we demonstrate that in smooth two-dimensional chaotic flows, mixing obeys a correlated aggregation process, where the spatial distribution of the number of lamellae in aggregates is highly correlated with their elongation and is set by the fractal properties of the advected material lines. We show that the presence of correlations makes mixing less efficient than a completely random aggregation process because lamellae with similar elongations and scalar levels tend to remain isolated from each other. We show that correlated aggregation is uniquely determined by a single exponent which quantifies the effective number of random aggregation events. These findings expand aggregation theories to a larger class of systems, which have relevance to various fundamental and applied mixing problems.
\end{abstract}
%
%\begin{keywords}
%\end{keywords}

%%%%%%%%%%%%%%%%%%%%%%%%%%%%%%%%%%%%%%%%%%%%%%%%%%%%%%%%%%%%
\section{Introduction} 
%%%%%%%%%%%%%%%%%%%%%%%%%%%%%%%%%%%%%%%%%%%%%%%%%%%%%%%%%%%%
Mixing of solutes by the stirring action of a fluid flow is ubiquitous to many natural and industrial processes \citep{OttinoARFM1990}.
The evolution of the solute concentration $c$ in an incompressible velocity field $\boldsymbol{v}$ is governed by the conservation equation
\begin{equation}
\partial_t c + \boldsymbol{v} \cdot \boldsymbol{\nabla}c = \kappa \boldsymbol{\nabla}^2  c,
\label{eq:AD}
\end{equation} 
with $\kappa$ the molecular diffusivity. While advected by the flow, the solute also \textit{mixes} with its surrounding due to the irreversible effect of molecular diffusion, and concentration tends to homogenize.

Although fully linear, Eq.~\ref{eq:AD} is characterized by rich spatio-temporal behaviors, that have retained considerable attention along the years~\citep{warhaft2000passive,falkovich2001particles,rothstein1999persistent}. As illustrated in Fig.~\ref{fig:intro}a, an initial blob of passive solute repeatedly stirred in a chaotic flow produces scalar filaments that stretch exponentially in time~\citep{Aref_1984}. Their thickness is blocked at the so-called Batchelor scale, scale at which molecular diffusion compensates the effect of stretching on scalar gradients~\citep{Batchelor1959}. Owing to flow incompressibility, these filaments are also exponentially compressed onto each other, leading to their diffusive coalescence (Fig.~\ref{fig:intro}b). These three processes --stretching, diffusion and coalescence-- are key to quantify and predict the statistics of $c$ in the homogenizing mixture.

Filament stretching and diffusion can be effectively quantified in a Lagrangian frame aligned with the local directions of elongation and compression of the flow field~\citep{Ranz1979,balkovsky1999universal}. Such approach, referred to as Lagrangian stretching or lamellar theory, provides a theoretical link between the stretching statistics of the flow (its advective properties) and the evolution of concentration in isolated lamellae. It yields good estimates of the decay of the scalar concentration variance~\citep{haynes2005controls,tsang2005exponential} and of the shape of the scalar pdf in a range of chaotic flows~\citep{Meunier2010}. However, it is \textit{a-priori} limited to the early stage of mixing, when scalar filaments evolve far from each-other, that is well before coalescence~\citep{fereday2002scalar,villermaux2003mixing}.  

Interestingly, the Lagrangian stretching framework has been shown~\citep{haynes2005controls,tsang2005exponential} to provide correct mixing rate predictions even at asymptotic time, for smooth flow in the so-called Batchelor regime, that is when scalar gradients persist only at small scales. Indeed, in periodic domains, asymptotic mixing rates have been shown to be either controlled "locally" by the stretching of material lines, or "globally" by macro-dispersive flow properties, depending on the domain size with respect to the velocity correlation length. In the "local" case, it is still unknown why the Lagrangian stretching framework would remain accurate after coalescence time, when stretched filaments are not isolated from each other. In addition, it is unknown how stretching statistics controls the shape of the asymptotic scalar pdf. Some theoretical results exist on the limiting shape of scalar pdf~\citep{sinai1989limiting}, but owing to closures that are not trivial to relate to physical properties of the flow~\citep{sukhatme2004probability}.  In turn, \citet{pierrehumbert1994tracer} and followers~\cite{rothstein1999persistent,sukhatme2002decay} have documented the self-similarity of asymptotic scalar pdf through the emergence of a ``strange eigenmode'', but a physically based modeling of the shape of the scalar pdf is still an open question. % strange eigenmode

Recognizing the additive nature of scalar coalescence in bounded domains, \citet{villermaux2003mixing} have proposed to describe the late time evolution of concentration pdf by a random aggregation mechanism, whereby the concentration field results from a sum of independent filaments that individually obey Lagrangian stretching dynamics, and that overlap by diffusion. This random aggregation scenario accurately captures the shape of asymptotic scalar pdf ~\citep{Meunier2010,DuplatVillermaux08}, if one uses the rate of filament aggregation as a fitting parameter. However, no independent measure of filament aggregation in chaotic flows has been obtained so far to validate the random aggregation hypothesis. In addition, it is still unclear if aggregation processes tends to accelerate or decelerate mixing compared to stretching.

Recently, \citet{HeymanPRL2020} have proposed that scalar aggregation could obey a correlated process, where poorly stretched regions of the flow containing weakly mixed solutes remain weakly aggregated in proportion. Such intimate correlation between stretching and aggregation has the advantage to explain the persistent role of Lagrangian stretching statistics on asymptotic scalar mixing which is largely observed in the Batchelor regime~\citep{tsang2005exponential}.

Here, we develop a general aggregation theory for scalar mixing in smooth, two-dimensional, and fully chaotic flows in the Batchelor regime. The theory is based on the description of the spatial clustering of advected material lines, and its correlation with local elongation rates.  The theory allows describing the asymptotic evolution of the scalar pdf and its moments at any Péclet number based on the sole knowledge of Lagrangian stretching statistics and fractal dimensions of the advective material lines. It may be generalized to a wider range of flows and boundary conditions.

The paper is organized as follows: in the first section, we recall important concepts and variables pertaining to Lagrangian stretching and aggregation frameworks. In the second section, we present the two-dimensional incompressible chaotic flows used to illustrate and validate the general aggregation theory, and their numerical resolution. In the third section, we focus on the geometry of material lines advected by a chaotic flow and we derive the evolution of the pdf of the number of advective filaments in aggregates, as well as the statistics of elongation in aggregates. In section 4, we describe how the scalar concentration pdf and its moment can be obtained from the knowledge of both the distribution of aggregate size and filaments statistics in aggregates. 
%
%discuss the two main hypothesis proposed to describe lamella aggregation in heterogeneous flows (section 2). We then use chaotic flow simulations to derive a new correlated aggregation theory. In Section 3, we describe the fractal feature of material lines in heterogeneous chaotic flows and its link to the distribution of the number of aggregated lamellae. In Section 4, we investigate the properties of correlated aggregation. In Section 5, we derive a model for the aggregated scalar pdf.  
%
%The paper is organised as follows.

\section{Lagrangian stretching and aggregation}

\subsection{Stretching and dilution}
We focus on smooth, two-dimensional, incompressible and time-varying flows, for which there exists two Lyapunov exponents of opposite signs. Advection thus creates elongated one-dimensional structures, called lamella or filament, that are the backbone of scalar mixing. 

The temporal evolution of single scalar filaments is trackable in a Lagrangian frame with a coordinate system $(x,y)$ advected with the flow and aligned with the directions of compression ($x$) and elongation ($y$)~\citep{Ranz1979,Villermaux2019}.  Because of their elongated shape, the concentration of lamellae is approximately constant in the $y$ direction. Thus, $\partial_y c$ is negligible compared to $\partial_x c$, and the two-dimensional advection-diffusion problem \eqref{eq:AD} simplifies to a one-dimensional advection-diffusion equation
\begin{equation}
\partial_t c + u(x) \partial_x c = \kappa  \partial^2_x  c,
\label{eq:AD1D}
\end{equation} 
with $u=-x \lambda(t)$ the velocity at which solute particles are compressed in the direction $x$ and $\lambda(t) \geq 0$ the stretching rate.  In two-dimensional incompressible flows, the stretching rate $\lambda(t)$ in the $y$-direction leads to a compression rate $-\lambda(t)$ in the $x$-direction. In chaotic flows, the mean stretching rate $\mu_\lambda$ is positive and equal to the Lyapunov exponent. We define the total lamella elongation as
\begin{equation}
\rho(t)=\exp\left(\int_0^t \lambda(t') \text{d}t'\right).
\label{eq:rho}
\end{equation}
%This approximation is valid when the characteristic compression time $\gamma^{-1}$ is smaller than the characteristic diffusion time $s_0^2 / D$, where $s_0^2$ is the initial lamella width, which is for $\text{Pe}_0=\gamma s_0^2 / \kappa > 1$~\citep{Villermaux2019}.
\citet{Ranz1979} showed that Eq.~\eqref{eq:AD1D} transforms to the simple diffusion equation
\begin{equation}
\partial_\tau c =  \partial^2_\xi  c,
\label{eq:AD1D_simple}
\end{equation}
if time is rescaled by 
\begin{equation}
\tau(t)=\frac{\kappa}{s_0^2} \int_0^t \rho(t')^2 \text{d}t',
\label{eq:tau}
\end{equation}
and space is rescaled by $\xi=x \rho /s_0$,  with $s_0$ the initial lamella width. For a Gaussian initial condition $c(\xi,0)=\theta_0\exp(-\xi^2)$, the solution of  \eqref{eq:AD1D_simple} is
\begin{equation}
c(\xi,\tau)=\frac{\theta_0}{\sqrt{1+4 \tau}} \exp(-(\xi/\sqrt{1+4 \tau})^2).
\label{eq:gaussian}
\end{equation}
Equivalently in the original Lagrangian coordinate system $(x,y)$,
\begin{equation}
\Theta(x,t)=\theta(t)\exp(-(x/s(t))^2),
\label{eq:isolatedstripdistribution}
\end{equation}
where $\theta$ is the maximum concentration of the lamella, following
\begin{equation}
\theta(t)=\frac{\theta_0}{\sqrt{1+4 \tau(t)}},
\label{eq:theta0}
\end{equation}
and $s$ the unit lamella thickness, following
\begin{equation}
s(t)=\frac{s_0 \sqrt{1+4 \tau(t)} }{\rho(t)}.
\label{eq:s}
\end{equation}
%Note that the scalar mass in a given cross-section in the direction $x$,
%\begin{equation}
%m(t)=\sqrt{\pi} \theta s = \sqrt{\pi} \theta_0 s_0 \rho(t)^{-1},
%\label{eq:mass}
%\end{equation}
%depends only on the final elongation state $\rho(t)$. Indeed, multiplying by the lamella elongation recovers mass conservation.
Note that the scalar mass per unit length under a lamella simplifies to
\begin{equation}
	m(t) = \sqrt{\pi} \theta s(t) = \sqrt{\pi} s_0 \theta_0 \rho^{-1}(t).
	\label{eq:mass}
\end{equation}

Since in random flows, the stretching rate is a random variable of time, the elongation of lamellae and the rescaled time are also randomly distributed.  An approximation of the statistics of $\tau$ was proposed~\citep{Meunier2010,lester2016chaotic} as 
\begin{equation}
\tau \approx \frac{\kappa }{ 2 s_0^2} \frac{ t }{ \log \rho} (\rho^2-1),
\end{equation}
recognising the fact that the last stretching events have a predominant weight in the stochastic integral~\eqref{eq:tau}. 
At large time, $\rho\gg 1$ and $\log \rho/t \to  \mu_\lambda$. Note that the notation 
\begin{equation}
	\mu_{\bullet}=\langle \bullet	 \rangle
\end{equation}
is used thorough the paper for a mean quantity. Thus,
\begin{equation}
\tau \to  \frac{\kappa}{ 2 \mu_\lambda s_0^2} \rho^{2} = \frac{1}{4} \left( \frac{s_B}{s_0} \rho\right)^2,
\end{equation}
with the so-called Batchelor scale
\begin{equation}
	s_B=\sqrt{2 \kappa / \mu_\lambda}.
	\label{eq:sb}
\end{equation}
Note that, in his seminal paper, \citet{Batchelor1959} uses the mean turbulent strain rate instead of the Lyapunov exponent $\mu_\lambda$ to quantify fluid stretching.
Thus, filaments dilute at large times, with a peak concentration following 
\begin{equation}
\theta (t) \to  \frac{\theta_0 s_0}{ s_B} \rho^{-1}(t)
\label{eq:theta}
\end{equation}
while their thickness tends to the Batchelor scale $s \to s_B$. In contrast, the length of filaments grows as
\begin{equation}
	L(t) = \ell_0 \mu_\rho (t),
\end{equation}
where $\ell_0$ is the initial filament length. Eq.~\eqref{eq:theta} shows that the concentration of single diffusing filaments is inversely proportional to their elongation. Thus, describing the advective stretching statistics is sufficient to describe the mixing of passive scalar. Indeed, the spatial variance of an elongated material line in a flow domain of area $\mathcal{A}$ can be obtained by integration of the lamellar concentration (Eq.~\eqref{eq:theta}) squared along the filament path \citep{Meunier2010}, giving 
\begin{equation}
	\sigma_c^2 \sim \frac{1}{\mathcal{A}} L(t) \mu_\rho^{-2} = \frac{1}{\mathcal{A}} \mu_\rho^{-1}.
	\label{eq:isolated2}
\end{equation}

Assuming a normal distribution of stretching with mean $\mu_\lambda t$ and variance $\sigma^2_\lambda t$, the scalar variance decays can be estimated from the stretching statistics. For flows with moderate stretching variability, where $\sigma^2_\lambda\leq \mu_\lambda$,
\begin{equation}
		\sigma_c^2 \sim \exp\left(-(\mu_\lambda-\sigma_\lambda^2/2)t\right).
	\label{eq:variance_sol}
\end{equation}
In contrast, for flows with larger stretching variability where $\sigma^2_\lambda \geq \mu_\lambda$, the value of $\mu_\rho^{-1}$ is entirely controlled by small elongations with $\rho \leq 1$ ~\citep{balkovsky1999universal}. In that case, $\mu_\rho^{-1}\sim 	\exp(-\mu_\lambda^2/(2\sigma_\lambda^2)t).$

%Typical asymptotic scaling of $\mu_\lambda\rho^{-1}$ in the baker map and the sine flow are reported in Table.~\ref{tab:table1}. 

It is important to note that the Lagrangian stretching approximation is correct in the limit of one dimensional filaments, that is when the curvature radius is much larger than the diffusing filament width $s_B$. In bounded flow domains, there exist regions where high curvature develop~\citep{tang1996} which are known~\citep{Thiffeault2004} to be associated with weak stretching rates (low elongations) and thus low mixing (high concentrations). This questions the ability of the Lagrangian stretching framework to capture the tails of the scalar concentration pdf. However, at small $\kappa$ (small $s_B$), the impact of these regions on global mixing rates should remain limited. This is confirmed by the good agreement of the Lagrangian stretching theory with observed mixing rates~\citep{haynes2005controls}.

\subsection{Aggregation}

\begin{figure}
\centering \includegraphics[width=0.9\linewidth]{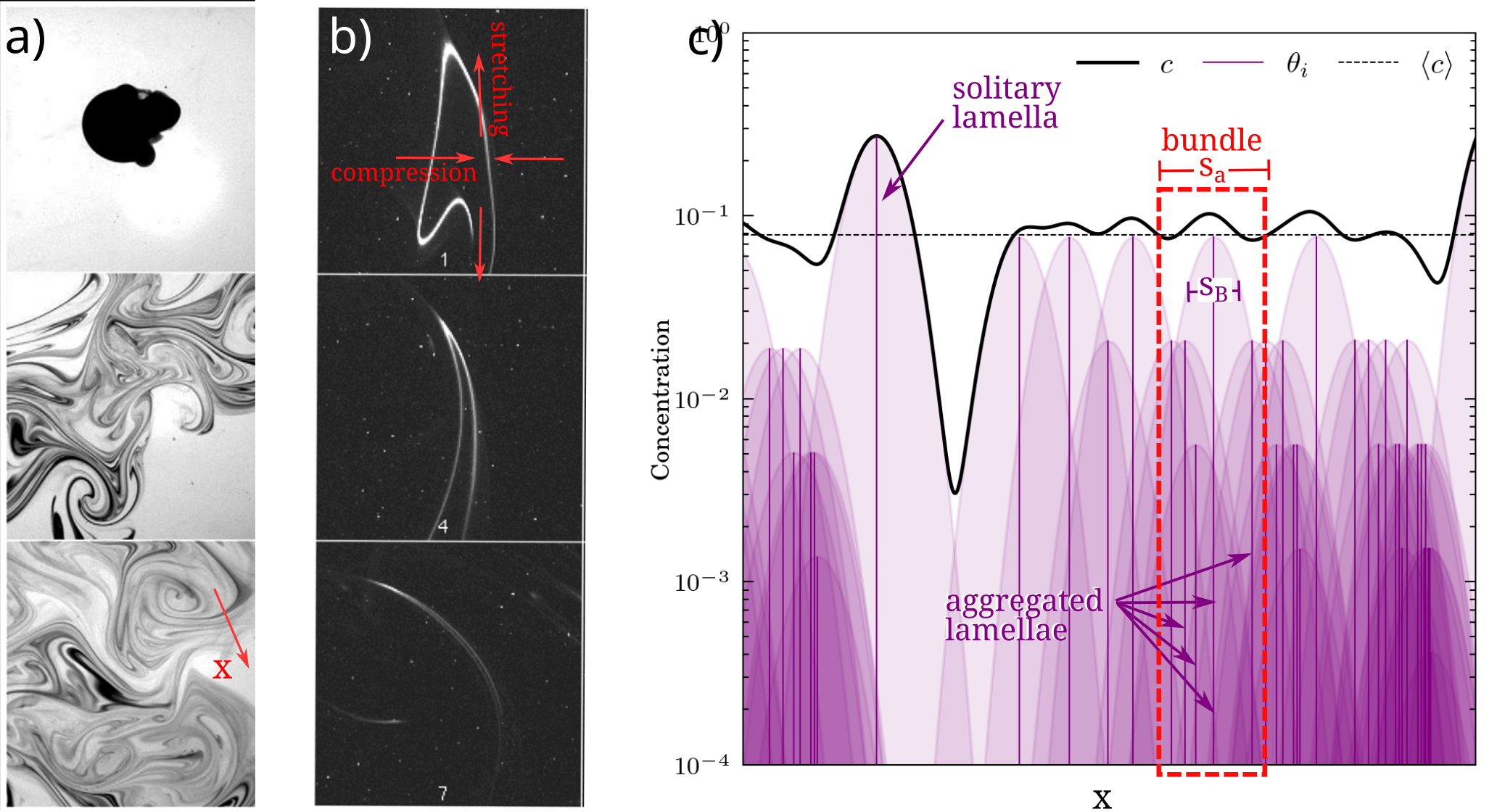}
\caption{a. Mixing of a diffusive scalar by a random stirring protocol (time sequence top to bottom), evidencing the apparition of stretched scalar filaments (adapted from~\citet{villermaux2012dissipation}). b. Blow up on the coalescence of neighbouring filaments under the action of compression (adapted from \citet{DuplatVillermaux08}) c. Concentration profile of a scalar field showing the coexistence of solitary filaments and bundles of filaments. The scalar concentration $c$ is obtained by the superposition of individual lamellae in a bundle of size $s_a$. All lamellae have a Gaussian shape with decaying maximum concentration $\theta_i$ (Eq.~\eqref{eq:theta}) and width tending to $s_B$.} \label{fig:intro}
\end{figure}

The production of curvature tends to produce folds in the material line (Fig.~\ref{fig:intro}b). The distance between separated parts of the elongated material line is thus reduced exponentially via fluid compression. This mechanism creates a highly foliated structure at late time (Fig.~\ref{fig:intro}a). Individual filaments are thus no longer isolated, and start to coalesce on length scales comparable to their width $s_B$. This so-called aggregation regime~\citep{villermaux2003mixing} has two properties. 

First, filaments tend to accumulate locally due to exponential flow compression. \textit{Bundles} of aggregated lamellae are thus formed by individual filaments sharing the same diffusive neighbourhood of size $\sim s_B$. The time at which aggregation initiates is when the total area of a lamella of length $L(t)$ and width $s_B$ is greater than the available area of the flow domain $\mathcal{A}$~\citep{garrett1983initial}, e.g.
\begin{equation}
L(t) s_B \gtrsim \mathcal{A},
\label{eq:lA}
\end{equation}
The mean number of lamella $n$ in each bundles is thus
\begin{equation}
\mu_{n}(t)  \sim \frac{L(t) s_B }{  \mathcal{A}}.
\label{eq:n_estimate}
\end{equation}
Assuming a constant stretching rate $\lambda$, the length of lamella is $L(t)  = \ell_0 \exp(\lambda  t)$, the coalescence time $t_c$ at which Eq.~\ref{eq:lA} is first fulfilled  is
\begin{equation}
t_c \sim \frac{1}{\lambda} \log\left(\frac{ \mathcal{A}}{\ell_0 s_B} \right).
\label{eq:coalescencetime}
\end{equation}

Second, the linearity of the advection-diffusion equation implies that scalar concentration fields can be decomposed into a sum over the concentration profiles of solitary lamellae~\citep{le2017scalar}. 
The scalar concentration $c$ at a given position can be constructed as the sum of concentrations $\theta_i$  of individual lamella $i$ (and defined by Eq.~\eqref{eq:theta}) present in a small neighborhood around this position, of size $s_a$ comparable to the asymptotical size of lamella, $s_B$, that is
\begin{equation}
c(t) \sim \sum_{i=1}^{n(t)} \theta_{i}(t)\sim \sum_{i=1}^{n(t)} \rho_{i}^{-1}(t),
\label{eq:sum}
\end{equation}
where the asymptotic relation between $\theta$ and $\rho$ (Eq.~\eqref{eq:theta}) was used.
%While each filament concentrations $\theta_{i}$ tends to 0 as large times (via Eq.~\eqref{eq:theta}), the sum \eqref{eq:sum} tends to a constant equal to the mean concentration $\mu_c$. 
Through Eq.~\eqref{eq:sum}, the aggregation process offers an appealing way to model the statistics of the scalar concentration field $c$ via the knowledge of the aggregated number of filaments $n$ at the aggregation+ scale $s_a$, their elongations $\rho_i$, and the possible correlations between these two variable.
  
%Two scenarii have been proposed to describe the statistical properties of the sum \eqref{eq:sum}: a fully random~\citep{DuplatVillermauxPRL2003} and a fully correlated~\citep{HeymanPRL2020} aggregation processes. These scenarii correspond to two caricatural routes towards homogeneity, described below.

%\section{Random vs correlated aggregation} 
%

%Random aggregation
%It was therefore assumed that the stirring action of turbulent flows is sufficiently random for the aggregation of individual filaments to be decoupled from their individual stretching histories.
Assuming no correlations between $n$ and $\rho_i$, and among $\rho_i$, leads to a \textit{random} aggregation scenario. The scalar concentration $c$ in a bundle is thus formed by the sum of independent and identically distributed random variables, following the solitary filament concentration pdf. The scalar concentration pdf, $P_c(c,t)$, thus results from the $n$-convolution of the isolated lamella concentration pdf $P_{1/\rho}(\rho^{-1},t)$, with the mean number of aggregations $\mu_n$ given by Eq.~\eqref{eq:n_estimate}~\citep{DuplatVillermaux08}. If $P_{1/\rho}$ is exponential or gamma distributed, then $P_c$ is a gamma distribution. Note that if $P_{1/\rho}$ is log-normal, the $n-$convolution pdf is not explicit, although its moments are given by the central limit theorem~\citep{schwartz1982distribution}. This random conjecture was shown~\citep{DuplatVillermaux08} to correctly describe mixing in turbulent flows, for which the velocity cascade over a large range of scales favor the decorrelation of filament stretching histories. 

Below the characteristic velocity length scale (the Batchelor regime), however, filament aggregation seems not to obey completely random dynamics. To see that, it is useful to consider the decay of scalar variance. In the random hypothesis, by the central limit theorem, scalar variance decays as the inverse of the aggregation number, that is, the inverse of the material length (Eq.~\eqref{eq:n_estimate}):
\begin{equation}
\sigma^2_c  = \frac{\mu_c^2}{\mu_n} \sim 1/ L(t).
\label{eq:random}
\end{equation}
Assuming a normal distribution of stretching rates,  $ L \sim \exp((\mu_\lambda+\sigma_\lambda^2/2)t)$ and thus the asymptotic scalar variance decay exponent is $\mu_\lambda+\sigma_\lambda^2/2$, larger than the decay exponent of solitary strips ($\mu_\lambda-\sigma_\lambda^2/2$, Eq.~\eqref{eq:variance_sol}). This result is in contradiction with observations suggesting the same decay exponent before and after aggregation time in the Batchelor regime~\citep{fereday2002scalar,tsang2005exponential}. 

We attribute this failure to the existence of correlations in the aggregation process below the characteristic velocity length scale. Indeed, in incompressible flows, lamella elongation is always balanced with transverse compression (Fig.~\ref{fig:intro}b), which attracts neighbouring lamella onto each-other. Thus, highly stretched lamella tend to be highly aggregated. Conversely, weakly stretched lamella have also experienced little compression, thus remaining isolated from the bulk and their evolution well described by the isolated lamellar theory. Since these lamella bears high concentration levels, they must dominate the statistics of scalar fluctuations at late time. Such correlations may explain why the Lagrangian stretching prediction for the scalar dissipation remains accurate long after aggregation time. 

Evidence of correlated aggregation was also observed experimentally during the chaotic mixing of two dye blobs~\citep{Duplat::PRL::2010}. If injected in a concentric manner the blobs locally experience similar stretching rates and aggregate in a correlated manner. In contrast, when placed at a certain distance from each other (larger than the characteristic length scale of the velocity field), their concentrations obey random aggregation rules. This observation suggests that stretching and aggregation are strongly correlated below the characteristic velocity length scale while they become uncorrelated above this scale. 
 
The remaining of the paper aims at establishing the laws governing correlated aggregation for scalar mixing by smooth chaotic flows, and at deducing their impact on the evolution of scalar concentration pdf.

\section{Numerical simulation of synthetic chaotic flows} 
We focus on scalar mixing in the so-called Batchelor regime~\citep{haynes2005controls} where the velocity field is smooth and present a single length scale, much greater than the typical scalar fluctuation scale. The domain size and the velocity length scale are of comparable magnitude, $\mathcal{L}\sim \mathcal{L}_v \sim 1$; the so-called ``local'' mixing regime described in \citet{haynes2005controls}. Thus, no scalar gradients can persist at scales larger than the velocity scale.

% Such regime is also qualified as ``smooth'' flows because velocity gradients remain relatively constant at the scale of $s_B$. This is in contrast to ``rough'' flows (e.g., turbulent flows at low Schmidt numbers) where the smaller flow scales lie below the Batchelor scale. 

\begin{figure}
\centering \includegraphics[width=\linewidth]{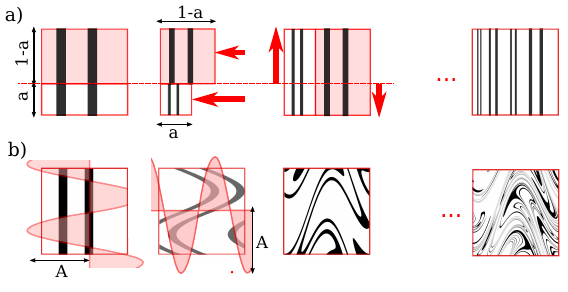}
\caption{ a) Transformations operated by the incompressible baker map with parameter $a$. First, the domain is cut horizontally at $y=a$, where $a$ is a constant between $0$ and $0.5$.  Uniform fluid compression operate on the domain parts $y<a$ and $y>a$ with  $a$ and $1-a$ respectively. Then, vertical stretching occurs with a factor $1-a$ and $a$ in these two regions, preserving the total area. b) Transformations operated by the sine flow with amplitude $A$. The flow is an alternation of horizontal and vertical sinusoidal velocity waves with amplitude $A$ and period $2\pi$. Random phases are chosen at each time periods so that the flow is fully chaotic.} \label{fig:baker&sine}
\end{figure}

To illustrate the geometrical features of aggregation, we use two synthetic chaotic transformations that complies with these conditions, namely the baker map and the random sine flow (Fig.~\ref{fig:baker&sine}). These are simple sequential advective maps that have been widely used to investigate the properties of chaos~\citep{finn1988chaotic,ott1989fractal,tsang2005exponential,giona2001geometry,Meunier2010,meunier2022diffuselet}.
We recall their definitions below.
%\jo{comment on global regime}

% Our objective is to uncover the kinematics of aggregation in these prototype flows. 

\subsection{Incompressible baker map}
The incompressible baker map~\citep{ott1989fractal,wonhas2002mixing} is a discontinuous transformation that operates on a two-dimensional periodic domain $[0,1]\times [0,1]$. The transformation writes
\begin{eqnarray*}
x_{t+1} &=& \left\lbrace \begin{array}{ll}
a x_t &\text{ if } y_t < a \\
1-(1-a) x_t &\text{ if } y_t > a \\
\end{array} \right. ,\\
y_{t+1}  &=& \left\lbrace \begin{array}{ll}
y_t / a &\text{ if } y_t < a \\
(1-y_t)/(1-a) &\text{ if } y_t > a \\
\end{array} \right. ,\\
\label{eq:baker}
\end{eqnarray*}
where $a \in [0,0.5]$ is a parameter controlling the heterogeneity of the map. A visual sketch of the map operation is plotted in (Fig.~\ref{fig:baker&sine}a).

An advantage of the baker map is that purely vertical scalar patterns (for which $c(x,y)=f(x)$) remain one-dimensional after application of the map, thus simplifying the problem to a single dimension. This simplicity allows for the analytical derivation of many features of the map, as we will show later. Another advantage is that it is possible to explore a wide range of stretching heterogeneity by varying $a$ between 0 and 0.5.  Indeed, the first two moments of stretching rate in the baker map are
\begin{eqnarray}
\mu_\lambda/ t &=& -a\log(a) - (1-a) \log(1-a)  ,\\
\sigma^2_\lambda / t &=& a(1-a)(\log(1-a)-\log(a))^2  .
\end{eqnarray}
Thus, for $a=0.01$, $\sigma_\lambda^2/\mu_\lambda= 3.7$  while for $a=0.49$, $\sigma_\lambda^2/\mu_\lambda=5.7 \cdot 10^{-4}$. It is important to note that this map involves discontinuous transformations, or ``cuts", that are absent in continuous flows such as turbulence but are common in flows through porous media~\citep{Lester2013}. This map also  prevents the formation of folds and cusps, where the one-dimensional hypothesis of the Lagrangian stretching framework is violated. The correspondence between Eq.~\ref{eq:AD} and Eq.~\ref{eq:AD1D} is thus exact in this map. 

\subsection{Random sine flow}
The random sine flow~\citep{pierrehumbert1994tracer,haynes2005controls} is a continuous transformation operating on a periodic domain $[0,1]\times [0,1]$ (Fig.~\ref{fig:baker&sine}b). The flow is periodic in time and space and reads for a given time period $t$
\begin{eqnarray*}
y_{t'+\delta t}  &=& y_{t'} +A\delta t \left\lbrace \begin{array}{ll}
\sin( 2\pi x_t' + \phi_t)  &\text{ for  } t<t'<t+1/2, \\
0 & \text{ for  } t+1/2<t'<t+1 \\
\end{array} \right. ,\\
x_{t'+\delta t} &=& x_{t'} +  A \delta t \left\lbrace \begin{array}{ll}
0 &\text{ for  } t<t'<t+1/2, \\
\sin( 2\pi y_t' + \psi_t)  & \text{ for  } t+1/2<t'<t+1 \\
\end{array} \right. ,\\
\label{eq:sineflow}
\end{eqnarray*}
%\begin{eqnarray}
%(x,y)_{t+1} &=& A (0, \sin( 2\pi x_{t} + \phi_m^x)) \text{ for } m<t<m+1/2, \\
%(x,y)_{t+1} &=& A (\sin( 2\pi y + \phi_m^y),0) \text{ for } m+1/2<t<m+1,
%\label{eq:sineflow}
%\end{eqnarray}
where the amplitude $A$ is a positive constant and $\phi_t,\psi_t$ are random phases that change at each time period $t$, and $\delta t$ the time step. The flow velocity having a single component, incompressibility is enforced. Scalar transport is continuous and considered on a periodic domain $[0,1]\times [0,1]$, ensuring a \textit{local} control of mixing rates~\citep{haynes2005controls}. 

As most random chaotic flows, the elongation of material lines in sine flows approximately follows a log-normal distribution with parameters $\mu_\lambda t$ and $\sigma^2_\lambda t$ that depend on the amplitude $A$. The stretching heterogeneity is much less variable than in the baker map, with ratio $\sigma^2_\lambda/\mu_\lambda$ ranging from $1$ when $A\to 0$ to  $\sigma^2_\lambda/\mu_\lambda\approx 0.6$ for $A=1.8$.  The stretching statistics of random sine flows can be found in \citet{meunier2022diffuselet}. In Appendix~\ref{sec:SM_average}, we recall useful results concerning the distribution of filament elongation and its moments in the sine flow and the baker map.

%In the following, we study the fractal geometry of advected material lines and their clustering in these chaotic flows. 
 %The statistics of $\tau$, $\theta$ and $s$ can be further derived from the statistics of $\rho$.

\subsection{Numerical methods}

\begin{figure}
\centering  \includegraphics[width=\linewidth]{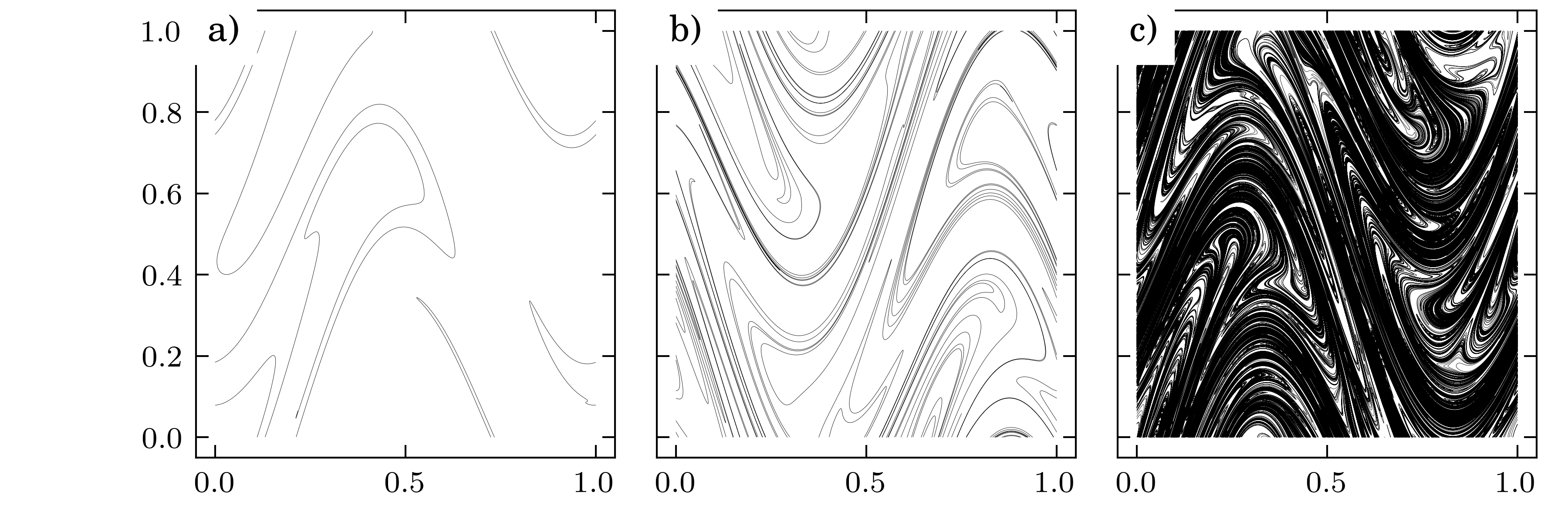}
\caption{ Advection of a material filament in the sine flow with $A=0.8$, for a) $t=3$ b) $t=5$ and c) $t=10$.} \label{fig:sinePDF}
\end{figure}

The hypothesis used through the paper is that the concentration field obeying Eq.~\eqref{eq:AD} is well approximated by a local summation over elementary lamellar concentrations, each of them individually following Eq.~\eqref{eq:AD1D}. This implies that the statistics of concentrations and their temporal evolution can in principle be inferred from the statistics of lamella elongation and aggregation. Thus, instead of directly solving the two-dimensional advection-diffusion equation (Eq~.\eqref{eq:AD}), our approach consists in computing elongation of a deforming Lagrangian support, advected by the flow (Fig.~\ref{fig:sinePDF}).% Local lamellar elongation is transformed into local lamellar concentration by Eq.~\eqref{eq:theta}, while the summation of neighbouring lamella concentrations results in the concentration field, approximated solution of Eq~.\eqref{eq:AD}.

Numerically, we advect a material line of initial length $\ell_0$ in the velocity field. In practice, the material line is defined by a series of particles $\boldsymbol{x}_i$ linked by segments. In the alternated random sine wave, the position of particles is tracked through time $t$ via an explicit Euler scheme (Fig.~\ref{fig:sinePDF}). The particles have constant velocities during a time step $\text{d}t =1/2$, so that this scheme is exact. The local elongation of segments is estimated as
\begin{equation}
\rho_i(t) = \rho_i(t-\text{d}t) \frac{\|\boldsymbol{x}_i(t+\text{d}t)-\boldsymbol{x}_{i+1}(t+\text{d}t)\|}{\|\boldsymbol{x}_i(t)-\boldsymbol{x}_{i+1}(t)\|}.
\end{equation}
The material line is refined after each time step to maintain a maximum local distance between points of $\text{d}x \leq 10^{-3}$. 

In the baker map, we start by a single filament aligned with the $y$ direction, so that we only need to track its coordinate $x$. Each operation  of the map doubles the number of filaments, with the first half being elongated by a factor $1/a$ and the second half by a factor $1/(1-a)$. We thus keep track of the growing number of filament positions $x_i$ and their elongations $\rho_i$.

The elongated and folded filament  is tracked up to the advection time where the total length is $L=10^7\ell_0$, limit corresponding to our computer memory. Local elongation of the advected material line can then be used to compute its local concentration, via Eq.~\eqref{eq:theta}. Aggregated scalar field can then be estimated by a local summation of individual lamellar concentration, via Eq.~\eqref{eq:sum}.

\subsection{Reconstruction of scalar fields}

As shown by \citet{Meunier2010}, it is possible to reconstruct the aggregated scalar field via the superposition of elementary lamella, represented as Gaussian ellipsoids of short axis $s$, that swipe along the advected material line. This reconstruction is exact in the Baker map since no cusps are forming. In the sine flow, it is only approximate in the regions of high curvature, where diffusion is genuinely two-dimensional. 

Here, we take an alternative reconstruction approach, which has the advantage of allowing analytical treatment. At late time, lamella widths tend to $s_B$ (Eq.~\eqref{eq:sb}), which is the minimum scale of fluctuations of the scalar field~\citep{Batchelor1959}. Thus, we construct the aggregated scalar field by binning lamella concentrations on a regular grid of size $s_a \sim s_B$.  
Consider a box of surface $s_a^2$ centred in position $\boldsymbol{x}$, the aggregated concentration level in this box can be constructed from the sum of the masses per unit length $m_i$ of the $n(\boldsymbol{x})$ individual lamellae of length $\ell_i$ present in this box
\begin{equation}
	c(\boldsymbol{x}) \approx    \frac{1}{s_a^2} \sum_{i=1}^{ n(\boldsymbol{x}) } s_a  m_i =   \frac{ \sqrt{\pi} \theta_0 s_0}{s_a} \sum_{i=1}^{n(\boldsymbol{x})}  \rho^{-1}_i.
	\label{eq:aggconcentration}
\end{equation} %We also provide in Appendix the main temporal scaling expected for elongation moments in the case of normally distributed stretching rates (e.g. the sine flow) and in the baker map.
where we used Eq.~\eqref{eq:mass} for the evolution of the solute mass per unit length carried by an individual lamella at a given location. Eq.~\eqref{eq:aggconcentration} forms the base of the statistical description of aggregation intended in this paper. 

In practice, we choose 
\begin{equation}
	s_a=\sqrt{2\pi} s_B,
	\label{eq:sA}
\end{equation}
for the variance of the regular grid reconstruction to match the variance of the true field (i.e., the one obtained by aggregating Gaussian lamellae). The factor $\sqrt{2\pi}$ is obtained so that both reconstruction methods provide the same mean-squared concentration in the case of a single lamella.

Note that the aggregated concentration field can be deduced at various Péclet numbers by varying the aggregation scale $s_a$ (Eq.~\eqref{eq:sA}), without recomputing the advective filament position. We define the Péclet number as the ratio of diffusive to stretching time scales, e.g.,
\begin{equation}
	\text{Pe}= \mu_\lambda \mathcal{L}_v^2/ \kappa.
\end{equation}
Through Eq.~\eqref{eq:sb}, $s_B= \mathcal{L}_v\sqrt{2/\text{Pe}}$ and thus $s_a= 2\sqrt{\pi}\mathcal{L}_v/\text{Pe}$.

This regular grid approximation of the aggregated scalar field disregards the exact Gaussian shape of lamellar concentrations and the variations of their width $s$ (Eq.~\eqref{eq:gaussian}).  However, classical box counting methods and fractal dimensions can be used to describe the geometrical and statistical features of the aggregated scalar field. Our goal is to describe the joint statistics of the variables $\rho_i$ and $n$ arising in Eq.~\eqref{eq:aggconcentration}, to infer the statistics of $c$.  %Thus, we forget^ the spatial dependency of $\rho$, $n$ and $c$, and consider these as random variables.

\begin{figure}
\includegraphics[width=\linewidth]{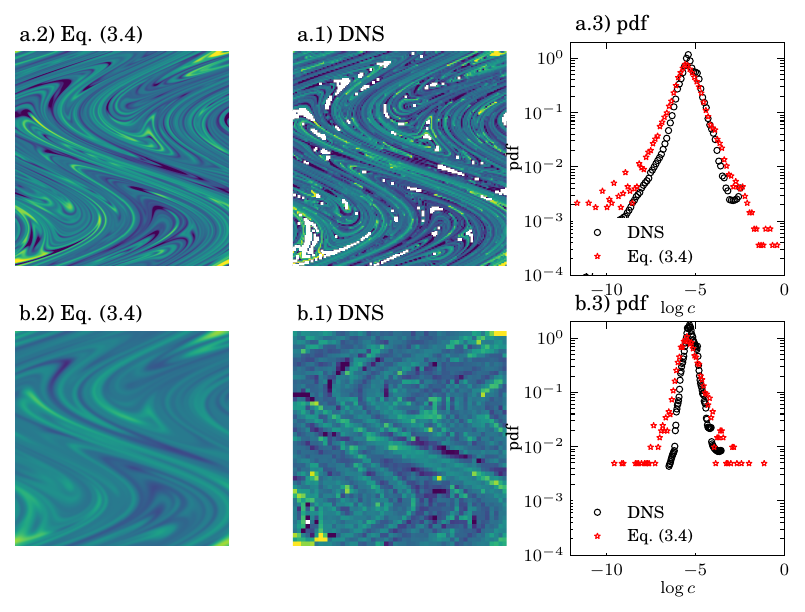}
\caption{Comparison of log-concentation fields obtained at time $t=10$ (fully aggregated regime) .1), with the aggregation framework ( Eq~\eqref{eq:aggconcentration}), .2) with Direct Numerical Simulation of Eq.~\eqref{eq:AD} and 3.) their pdf. The comparison is made for the sine flow ($A=0.8$) for two Péclet numbers, corresponding to $s_a=1/150$ (a.) and $s_a=1/50$ (b.)}\label{fig:quality}
\end{figure}

The approximation \eqref{eq:aggconcentration} is compared to direct numerical simulations (DNS) of the advection-diffusion equation for different Péclet numbers in Fig.~\ref{fig:quality}. DNS is obtained with the spectral method described in \citet{meunier2022diffuselet}, with a $2048^2$ grid and a time step of 0.1. There is a good qualitative match between the aggregated field and the DNS, match confirmed by comparing the distributions of concentrations. Thus, we conclude that the construction of the aggregated scalar field via Eq.~\eqref{eq:aggconcentration} is able to capture the essential statistical features of the two-dimensional advection-diffusion problem, and offers a convenient way to explore its statistics.

%% AGGREGATION
\section{Clustering properties of advected material lines}\label{sec:fractaln}
\begin{figure}
\centering \includegraphics[width=0.8\linewidth]{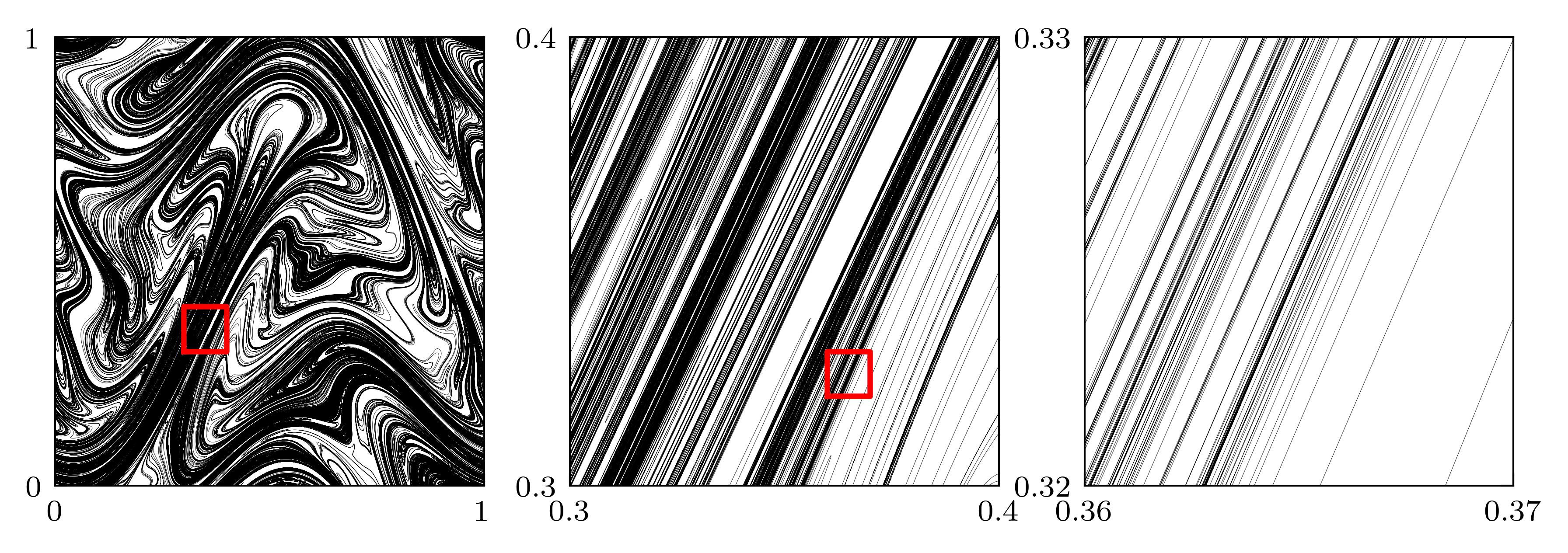}
\\ \centering   \includegraphics[width=0.8\linewidth]{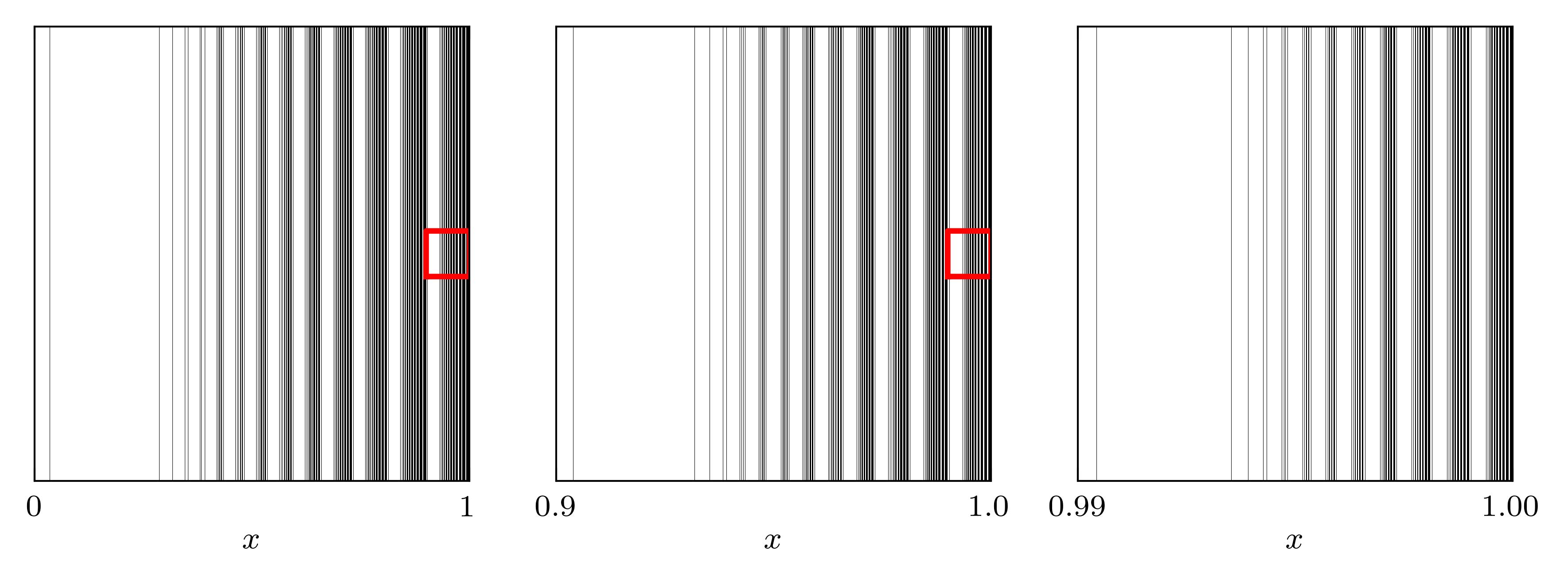}
\caption{Fractal geometry of material material lines in (top) the sine flow ($A=0.8$) and (bottom) the baker map ($a=0.1$) observed at different scales. The red square indicates the area selected for zooming.}\label{fig:geometry}
\end{figure}

In incompressible flows, the stretching of material elements by velocity gradients is compensated by transverse compression. Compression causes distances between lamellar elements to decrease exponentially over time. Smaller and smaller scales are thus continuously produced by flow compression.  Furthermore, in smooth chaotic flows, the typical scale of variation of velocity gradients is fixed and produces a heterogeneous stretching field for material lines. Dense (black) or diluted (white) regions of material lines are thus created  at large scale in the chaotic flow (Fig.~\ref{fig:geometry}). Such heterogeneous structures then cascade to smaller scales under the action of net compression, thus creating a fractal set of one-dimensional objects (lines) clustered around their transverse direction.  In two-dimensional incompressible flows, the Haussdorf dimension of this fractal set is necessarily  $D_0=2$, as per the Kaplan-York result~\citep{farmer1983dimension}.  Higher order dimensions can be smaller than 2 if stretching is heterogeneous.  To illustrate this, let us define a normalised measure $p_k$ with $k=1\cdots N$ defining a regular grid of bin size $\epsilon=\mathcal{L}/N$, where $\mathcal{L}$ is the domain size. For instance, $p_k$ may be defined as the local density of lamella in the bin, e.g. $p_k=n_k/n$ where n is the total number of lamella. %Since concentration levels of lamellae are additive, $p_k$ can be equivalently defined as the sum of lamella concentrations in one bin. 

The fractal dimension of order $q$ of the measure $p$ is obtained with~\citep{grassberger1983generalized}:
\begin{equation}
D_q -1 = \lim_{\epsilon \to 0} \frac{1}{q-1} \frac{ \log I_{q}(\epsilon)}{\log \epsilon}, \quad
I_q(\epsilon) \equiv \sum_k^{N=\mathcal{L}/\epsilon} p_k^q,
\end{equation}
where the subtraction of 1 on the left hand side accounts for the counting of one-dimensional structures (lamellae) in a two-dimensional domain.  This definition implies the following  spatial scaling  of the integral of the measure:
\begin{equation}
I_q(\epsilon) \sim \epsilon^{(q-1)(D_q-1)}.\label{eq:scalingIq}
\end{equation} 
In simple flows such as the baker map, $D_q$ can be obtained~\citep{finn1988chaotic} by observing the similarity properties of the map, which transfer at small scales the heterogeneity of the measure produced at large scales by a single operation of the map. Characterising the result of one elementary operation of map on the measure thus also informs about the spectrum of fractal dimensions. 
%\subsection{Fractal measure of the number of lamella}
%In the following, we derive this spectrum for the baker map (see also \citet{finn1988chaotic}) and show it is related to the distribution of $n$. 

We consider the measure of the local number of lamella in bin $k$, $p_k=n_k/n$. As shown in Fig.~\ref{fig:baker&sine}a, an operation of the baker map doubles the total number of these lamellae, while maintaining the same local distribution of lamellae on smaller bins of sizes $a\epsilon$ for $x<a$ and $(1-a)\epsilon$ for $x>a$. 
This similarity allows one to compute the first fractal dimension explicitly (see Appendix~\ref{app:baker} and \citet{finn1988chaotic}), 
 \begin{equation}
 D_0=2, \quad D_1 =1+ \frac{2 \log 2}{ \log (a^{-1} +(1-a)^{-1})}.
\label{eq:D1}
\end{equation}

\begin{figure}
\centering \includegraphics[width=7cm]{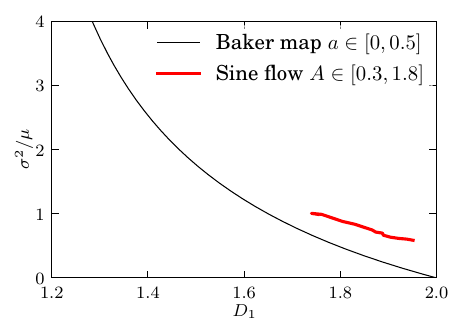}
\caption{Relation between stretching rate mean $\mu_\lambda$ and variance $\sigma_\lambda^2$ and fractal dimension $D_1$ in the baker map and sine flow with varying parameters $a$ and $A$.}\label{fig:fractal_stretch}
\end{figure}

For random flows such as the sine flow, \citet{ott1989fractal} argue that there exists a general relationship between stretching rate statistics and fractal dimensions as  $D_q =  f_q(\sigma_\lambda^2,\mu_\lambda)$, although a closed-form solution is not always as trivial as for the baker map. We show in Fig.~\ref{fig:fractal_stretch} that the ratio $\sigma_\lambda^2/\mu_\lambda$ is directly related to the fractal dimension $D_1$. This suggests that the fractal clustering of material lines is closely linked to the large-scale heterogeneity of stretching rates. Since the flow is smooth, the heterogeneity  created at large scales cascades to smaller scales, conserving its geometrical structure and creating a fractal geometry.

As suggested by Figure.~\ref{fig:fractal_stretch}, the function $f_1$ is different for baker map and the sine flow. Indeed, the ratio $\sigma_\lambda^2/\mu_\lambda$ tends to a positive constant in the sine flow when $A\to \infty$,  while $\sigma_\lambda^2/\mu_\lambda \to 0$ in the baker map when $a\to 0.5$. This finite limit comes from the fact that the sine flow is a continuous transformation with no cutting and thus does not tend to a uniform stretching rate. In the contrary, when $A\to0$, $\sigma_\lambda^2/\mu_\lambda \to 1$ which is a maximum bound for the ratio in the sine flow~\citep{meunier2022diffuselet}, thus limiting the possible range of fractal dimensions produced by continuous chaotic flows, compared to discontinuous maps.

\subsection{Spatial distribution of $n$}
\begin{figure}
\centering \includegraphics[width=10cm]{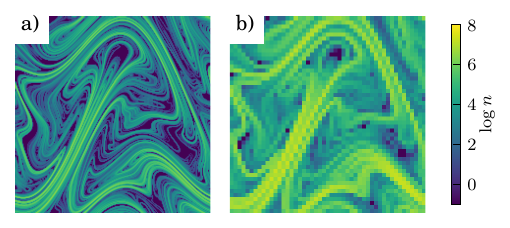}
\caption{Spatial distribution of the number $n$ of lamella in bundles defined by a regular grid of size a) $s_a=1/200$ and b) $s_a=1/50$, in the sine flow with parameter $A=0.5$.}\label{fig:sineN}
\end{figure}

\begin{figure}
\centering\includegraphics[width=0.5\linewidth]{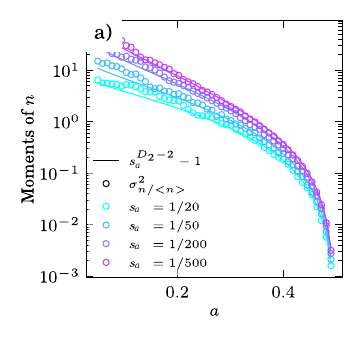}
\centering\includegraphics[width=0.46\linewidth]{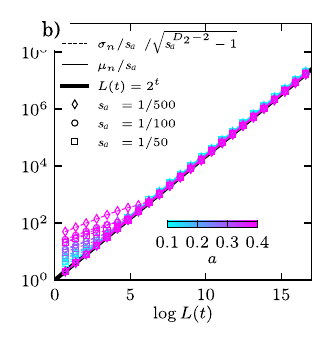}
\caption{a) Scaling of the spatial variance of $\log n$ as a function of $a$ in the baker map and theoretical prediction, Eq.~\eqref{eq:sigman_n}. b) First two moments of $P_n$ through time compared to theoretical predictions, Eqs.~\eqref{eq:mun} and \eqref{eq:sigman} in the baker map  ($\mathcal{A}=1$).}
\label{fig:momentNbaker}
\end{figure}

\begin{figure}
\centering\includegraphics[width=6cm]{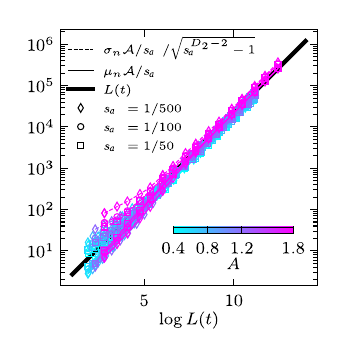}
\caption{First two moments of $P_n$ through time compared to theoretical predictions, Eqs.~\eqref{eq:mun} and \eqref{eq:sigman} in the sine flow ($\mathcal{A}=1$).}
\label{fig:momentNsine}
\end{figure}

The spatial distribution of the number $n$ of lamella per bin size $\mathcal{A}$ (Fig.~\ref{fig:sineN}) can be obtained as follows.  Comparing the mean area occupied by a filament of length $L(t)$ and aggregation scale $s_a$ to the domain surface $\mathcal{A}$, we get an estimate of the mean number of lamellae $\mu_n$ in bundles
\begin{equation}
\mu_n \sim L(t) s_a / \mathcal{A} .
\label{eq:mun}
\end{equation}
Higher moments can be obtained from a study of the fractal structure of material lines. To this end, we consider the spatial measure corresponding to the local number of lamellae in each bundle:
\begin{equation}
p_k \equiv \frac{n_k}{\sum_k n_k}
\label{eq:measure}
\end{equation}
The Renyi definition~\citep{grassberger1983generalized} of the fractal dimension of order 2 of this measure is
\begin{equation}
 D_2 -1  \approx \frac{ \log \sum_k p_k ^2}{ \log s_a},
\end{equation}
assuming $s_a\to 0$. Replacing Eq.~\eqref{eq:measure} in the last expression provides 
\begin{equation}
\sum_k \left( \frac{ n_k}{\sum_k n_k} \right) ^2= s_a^{D_2-1}.
\end{equation}
Since $\sum_k n_k= N \mu_n$ we have 
\begin{equation}
\sum_i \left( \frac{ n_k}{\sum_k n_k} \right)^2 = \frac{1}{N} \left\langle  \left({n}/{\mu_n}\right)^2 \right\rangle,
\end{equation}
with $N \approx \sqrt{\mathcal{A}}/s_a$ the number of bundles in the flow domain.
Since 
\begin{equation}
\sigma^2_{n/\mu_n} = \left\langle  \left(n/\mu_n\right)^2 \right\rangle- \left\langle n/\mu_n \right\rangle^2,
\end{equation}
then,
\begin{equation}
\sigma^2_{n/\mu_n} = \sqrt{\mathcal{A}} s_a^{D_2-2} - 1.
\label{eq:sigman_n}
\end{equation}
Thus the variance of $n/\mu_n$ reaches a constant at asymptotic times, which is given by the fractal dimension of order 2.
The spatial variance of $n$ is then 
\begin{equation}
\sigma^2_{n} = \mu_n^2 \left( \sqrt{\mathcal{A}} s_a^{D_2-2} - 1 \right)
\label{eq:sigman}
\end{equation}
The predictions of Eqs.~\eqref{eq:mun}-\eqref{eq:sigman} are plotted against time in Fig.~\ref{fig:momentNbaker} and Fig.~\ref{fig:momentNsine} showing good agreement with simulations for a large range of Batchelor scales $s_a$ and flow heterogeneity, characterized by the parameters $a$ for the Baker map and $A$ for the sine flow. Higher moments of $n$ can be obtained with similar scaling arguments and linked to fractal dimension of higher order. This is however out of the scope of the present study.

\begin{figure}
\includegraphics[width=0.32\linewidth]{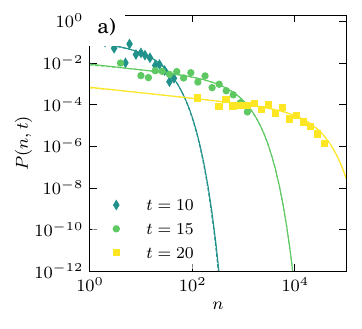}
\includegraphics[width=0.32\linewidth]{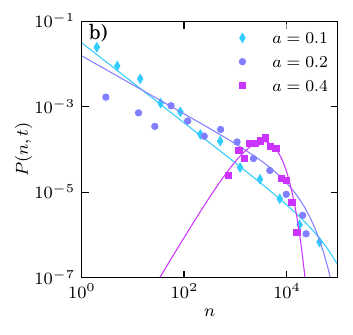}
\includegraphics[width=0.32\linewidth]{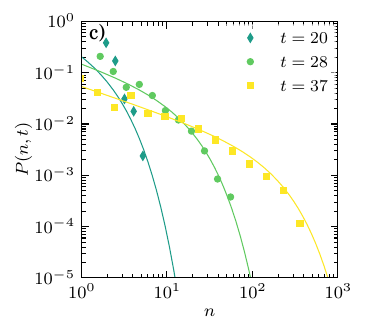}
\caption{$P(n,t)$ for the baker map and sine flow for $s_a=1/100$. Solid lines stand for the gamma pdf with theoretical moments given by Eq.~\eqref{eq:parameters_n} and dots stand for numerical simulations. a) baker map $a=0.3$ and variable $t$. b) baker map for $t=20$ and variable $a$.  c)  sine flow for $s_a=1/100$ and $A=0.4$.}\label{fig:pdfn_bakersine}
\end{figure}

Instead, we observed that the pdf of lamella aggregation number $P_n (n)$ is well fitted by a Gamma distribution
 (Fig.~\ref{fig:pdfn_bakersine}):
\begin{equation}
P_n (n) = \frac{1}{\Gamma(k_n) \theta^k_n} n^{k_n-1} \exp(-n/\theta_n),
\end{equation}
with $n \geq 0$ and $k_n,\theta_n$  defined by the moments of the distribution of n :
\begin{eqnarray}
k_n &=& \left( \sqrt{\mathcal{A}} s_a ^{D_2-2} -1 \right)^{-1},
\label{eq:kn} \\
\theta_n &=& \mu_n(t)\left( \sqrt{\mathcal{A}}  s_a ^{D_2-2} -1 \right),
\label{eq:parameters_n}
\end{eqnarray} 
with $\mu_n=L(t) \mathcal{A}/s_a$.  
%\begin{figure}
%\centering \includegraphics[width=7cm]{../Figures/scaling_k.pdf}
%\caption{Variation of power law exponent $n^{k_n-1}$ of $P_n$ with $D_1$ and $s_a$}\label{fig:pdfn_pw}
%\end{figure}
Note that the gamma distribution yields a power law distribution at small $n$ with exponent $k_n-1$ going from -1 to infinity with increasing $D_2$. For small $D_2$ (strong stretching heterogeneity), a significant part of the probability is concentrated at small $n$, thus in non-aggregated regions of the flow. We will show later that this can affects the value of negative moments of $n$.

%\section{Lamellar concentrations in bundles}
\begin{figure}
\centering  \includegraphics[width=0.7\linewidth]{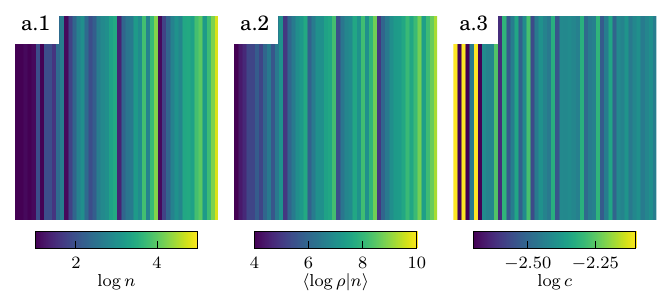}
\centering  \includegraphics[width=0.7\linewidth]{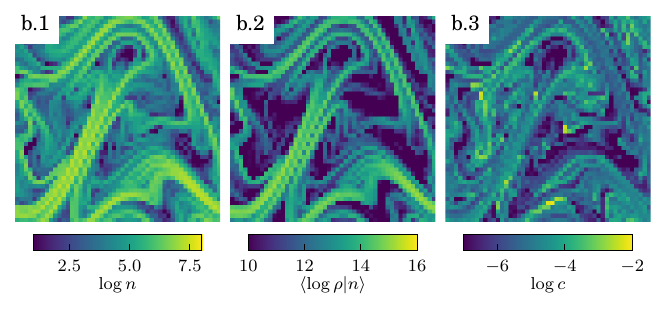}
\caption{Simulation of aggregation statistics in baker map ($a=0.3$)  and sine flow ($A=0.5$) for $s_a=1/50$~: .1 number of lamella $n$ in bundles, .2 mean of log-elongation in bundles and .3 sum of lamellar concentrations in bundles.}\label{fig:baker_nlag}
\end{figure}

\subsection{Local correlations between $n$ and $\rho$}
\begin{figure}
\centering \includegraphics[width=8cm]{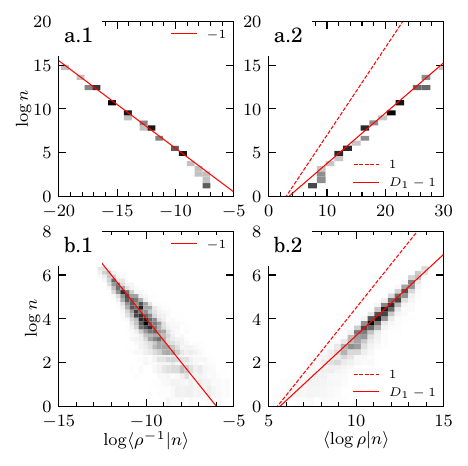}
\caption{Joint pdf (gray scale) of the number of lamellae in a bundle of size $s_a=1/200$ and (1) their mean inverse elongation (2) and their mean log-elongation for (a.) baker map ($a=0.1, t=24, D_1=1.57$) and (b.) sine flow ($A=0.8$, $t=10$, $ D_1=1.74$). The theoretical scaling of the measure (1) and (2), given by Eq.~\eqref{eq:nrho} and \eqref{eq:n:D1} respectively, are plotted as a continuous red lines with the slope indicated in the legend. Dashed red lines are guides to the eyes.}\label{fig:jointpdfbundle}
\end{figure}

Having precised the statistics of the bundle size $n$, we now explore the local statistics of lamella elongations in these bundles. To this end, we define the conditional averaging operator acting in lamellae located in the local neighbourhood of size $s_a$  by
\begin{equation}
\mu _{X | n } = \frac{1}{n} \sum_{i=1}^{n} X_i,
\end{equation}
where $X$ is a Lagrangian variable transported by lamellae and $n$ is the number of lamellae aggregated in the bundle. The remainder of this section is dedicated to uncovering the behaviour of conditional moments of elongation conditioned to $n$ (e.g. $\mu_{\rho^{-q} | n }$). Section~\ref{sec:unconditional} will then be dedicated to deriving unconditional probabilities by averaging on the distribution of $n$. %We show in Appendix~\ref{app:fractal} that the aggregated concentration levels $c$ mainly show scale independence, precluding a description the scalar moments of $c$ as the ones of $n$ the fractal geometry created by the chaotic flow. 

We plot in Fig.~\ref{fig:jointpdfbundle} the joint probability $P(n, \mu_{X | n} )$ obtained in the baker map and the sine flow for $X=\rho^{-1}$ and $X=\log \rho$, the inverse of elongation and the log-elongation of lamella respectively. Fig.~\ref{fig:jointpdfbundle} suggests that the following scaling holds in both flows:
\begin{equation}
\log n \sim - \log\mu_{\rho^{-1} | n} ,
\label{eq:nrho}
\end{equation}
which confirms the strong correlation between the number of lamellae in aggregates and their elongation.
%We have seen that The maximum concentration of lagrangian lamella follows
%\begin{equation}
%\theta(\tau)=\frac{\theta_0}{\sqrt{1+4\tau}} \approx  \frac{\theta_0 s_0}{s_a} \rho^{-1} \text{ for } \tau \gg 1, 
%\end{equation}
%where $s_a=\sqrt{\kappa/\lambda}$. The last approximation assumes $\tau \approx s_a^2/s_0^2(\rho^2-1)$, an assumption justified by the exponential nature of stretching in chaotic flows, which procures to the last stretching event a dominating role in the whole deformation history \citep{Meunier2010,Lester2016}.
Indeed, for large time, $c$ must tend to the conserved average scalar concentration $c \to \mu_c$. Thus, owing to Eq.~\eqref{eq:aggconcentration}, we must have 
\begin{equation}
n \sim 1/\mu_{\rho^{-1} | n },
\label{eq:cmaxN}
\end{equation}
a scaling that we confirm numerically (Fig.~\ref{fig:jointpdfbundle}). % In incompressible flows, the distance between lamellae $d$ is proportional to the amount of compression they have experienced, $\rho_i^{-1}$. Since the number of lamellae in a box of size $r$ is $n \sim 1 / \langle d_i | n \rangle$, then, $ n \sim 1/ \langle \rho^{-1} | n \rangle$, which recovers the above result. 

Fig.~\ref{fig:jointpdfbundle} also suggests that
\begin{equation}
\log n \sim (D_1-1) \mu_{\log \rho | n },
\label{eq:n:D1}
\end{equation}
where $D_1$ is the information dimension~\citep{ott1989fractal} of the measure $n$, and is given by Eq.~\eqref{eq:D1} for the baker map. Equation \eqref{eq:n:D1} can be derived analytically in the case of the baker map. Indeed, by the action of the map, the total number of lamellae increases  as $\log n = t \log 2$ while the mean log-elongation of these lamellae is $\mu_{\log \rho} =t ( -\log a - \log(1-a) )/ 2 $ leading to a constant ratio 
\begin{equation}
\frac{\log n}{\mu_{\log \rho }}= \frac{2\log 2}{\log a + \log(1-a)},
\end{equation}
 which is exactly the value of $D_1-1$ (Eq.~\eqref{eq:D1}). Assuming that the partition between $\log n$ and $\mu_{\log \rho}$ is preserved at small scales in each bundle, we have
\begin{equation}
\log n =  ({D_1 - 1})(  \mu_{\log {\rho} | n} -  \mu_{\log \rho_c}),
\label{eq:entropyN}
\end{equation}
with $ \mu_{\log \rho_c}$ a constant standing for the mean elongation at coalescence time.

The constant $\mu_{\log \rho_c}$ can be estimated by comparing the surface covered by the filament at the aggregation scale, $\mathcal{S}=s_a \rho_c L_0 $, with the domain area $\mathcal{A}$. The first aggregation event occurs when $\mathcal{S} \approx  \mathcal{A}$, that is when the elongation is $\rho_c \approx \mathcal{A}/(s_a \ell_0) $.  Eq.~\eqref{eq:entropyN} is indeed verified for baker map and sine flow with various parameters $a$ and $ \mathcal{A}$, with $\mu_{\log \rho_c} = - \log s_a$.

\subsection{Distribution of $\log \rho$ in a bundle of size $n$}
The two scaling laws $n \sim \mu_{\rho^{-1}| n}$ and  $\log n\sim (D_1-1) \mu_{\log \rho| n}$ provide key information about the heterogeneity of lamella elongations inside bundles.
Since the ensemble distribution of elongation $P_\rho(\rho)$ has a log-normal shape, we assume that the distribution of elongations inside bundles, denoted $P_{\rho|n}$, is also log-normally distributed. This implies that $\log \rho$ is normally distributed in bundles, with a mean
\begin{equation}
\mu_{\log \rho|n} \sim (D_1-1)^{-1} \log n.
 \label{eq:mu}
\end{equation}
 Since  $\log \mu_{\rho^{-1} | n}= -\mu_{\log \rho|n }+\sigma^2_{\log \rho |n}/2 \sim -\log n(\boldsymbol{x})$, the variance of log-elongation in bundles at large $n$ must be
\begin{equation}
\sigma^2_{\log \rho|n} \sim \frac{2(2-D_1)}{D_1-1} \log n. 
 \label{eq:sigma2}
\end{equation}
\begin{figure}
\centering \includegraphics[width=12cm]{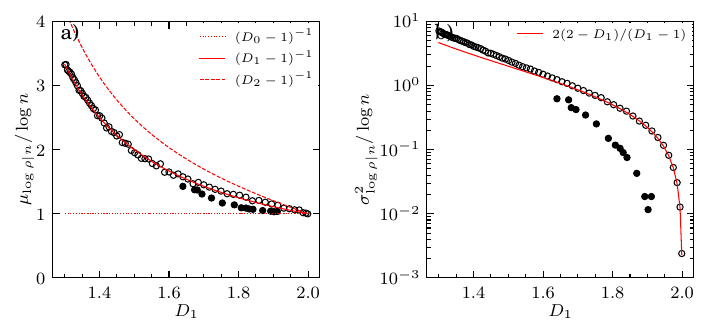}
\caption{Scaling of the mean (a) and variance (b) of log-elongation in bundles as a function of the information dimension $D_1$. Circles stands for numerical simulations in baker maps (open circles) and sine flow (filled circles). Continuous lines stands for theoretical prediction of the mean (Eq.~\eqref{eq:mu}) and variance (Eq.~\eqref{eq:sigma2}). Dashed lines are plotted to compare the mean with fractal dimension of other order. }\label{fig:musigma}
\end{figure}

We report in Fig.~\ref{fig:musigma} the simulated scaling $\mu_{\log \rho|n}/\log n$ and  $\sigma^2_{\log \rho|n}/\log n$ obtained asymptotically at large times. When $D_1\to 2$, $\mu_{\log \rho|n}/\log n \to 1$ while $\sigma^2_{\log \rho|n}/\log n \to 0$, meaning that bundles are formed by lamella of identical elongations.  In contrast, when $D_1 \to 1$, both $\mu_{\log \rho|n}/\log n $ and  $\sigma^2_{B}/\log n $ become infinite, while their ratio $\sigma^2_{\log \rho|n}/\mu_{\log \rho|n} = 2(2-D_1) \to 0.5 $. This limit suggests that the aggregation of lamellae remains correlated to their average elongation, although a fixed amount of stretching variability arises in bundles. A good agreement is found between theoretical prediction (Eqs.~\eqref{eq:mu}-\eqref{eq:sigma2}) and numerical simulations of aggregation in the baker map (Fig.~\ref{fig:musigma}). In contrast, the theory captures only qualitatively the behaviour of the random sine flow. This may be due to the continuity of the sine flow which produces curved lamellar structures whose dimension is not exactly one-dimensional.

%These results further invalidate the fully correlated aggregation hypothesis that assumes a uniform elongation in each bundle.  
The stretching variability in bundles is directly linked to the heterogeneity of the chaotic flow, because of the intimate relationship existing between the fractal geometry of the material line and the stretching statistics of fluid elements~\citep{ott1989fractal}. As such, it is impossible to have a single stretching rate per bundle as soon as the chaotic flow is heterogeneous and exhibits a distribution of stretching rates.  The absence of stretching variability in bundles ($\sigma^2_{\log \rho|n}=0$)  implies the uniformity of stretching at large scale ($\sigma^2_{\rho}=0$). This uniform case is reached when $D_1\to D_0=2$, for instance, in the baker map when $a \to 0.5$. In continuous flow maps such as the sine flow, regions of high and low stretching always coexist and $\sigma^2_{\log \rho|n}>0$.

\subsection{Moments of $1/\rho$ in a  bundle of size $n$}

\begin{figure}
\centering \includegraphics[width=\linewidth]{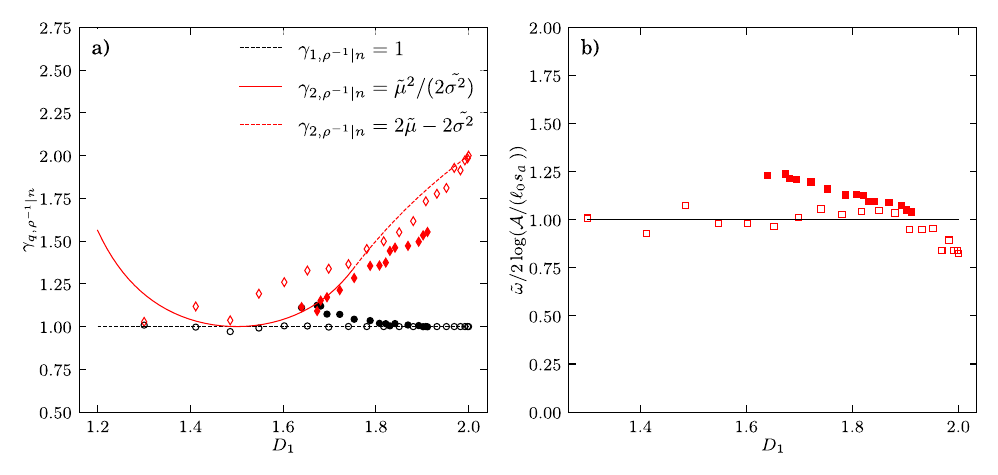}
\caption{a) Scaling exponents of the $q$ lamellar concentration moments in bundles (Eq.~\eqref{eq:scalingexponents}). Numerical estimates are plotted with symbols, red diamonds for $q=2$ and black circles for $q=1$. Unfilled and filled symbols represents simulations in baker map and sine flow respectively. Theoretical predictions (Eq.~\eqref{eq:tildegamma}) are represented by lines. b) Intercept $\tilde\omega$ of the scaling exponent of the $q$ lamellar concentration moments in bundles (Eq.~\eqref{eq:scalingexponents}) in baker map (empty squares) and sine flow (filled squares) and theoretical prediction (line, Eq.~\eqref{eq:omega2}).}\label{fig:scaling_theta}
\end{figure}

\begin{figure}
 \centering \includegraphics[width=7cm]{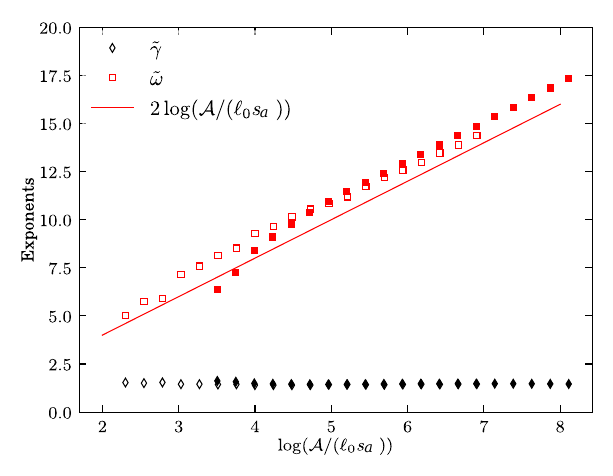}\caption{Dependence of $\tilde\gamma$ and $\tilde\omega$ with the aggregation scale in simulations (dots) of the baker map (empty symbols $a=0.2$) and the sine flow (filled symbols $A=1.2$) and comparison to theoretical prediction (Eq.~\eqref{eq:omega2}.}\label{fig:gamma2_omega2}
\end{figure}

Having described the first two moments of the distribution of lamella elongation in bundles  (Eqs~\eqref{eq:mu}-\eqref{eq:sigma2}), we now assume that the distribution is of log-normal shape. This choice is justified by the fact that elongation is a multiplicative process, thus usually leading to lognormal distributions~\citep{LeBorgneJFM2015, Souzy:2020aa}.  This allows us to compute the scaling of the $q-$moments of lamella concentrations in bundles, $\theta|n$. Owing to Eq.~\eqref{eq:theta}, we have
\begin{eqnarray}
\mu_{\theta^q | n} & \sim & \mu_{ \rho^{-q} | n }  \nonumber \\
&=& \int_1^\infty \rho^{-q} P_{\rho|n}(\rho) \text{d}\rho \nonumber \\ 
&\approx & \int_1^\infty e^{-(\log\rho-\mu_{\log \rho|n})^2/(2\sigma_{\log \rho|n}^2) - q \log \rho} \text{d}\rho.
\label{eq:scalingexponents}
\end{eqnarray}
The minimum bound for the integral is taken at $\rho=1$ and not 0, taking into account the fact that lamellar structures cannot be compressed in their longitudinal direction. As a consequence, $P_{\rho|n}(\rho)$ is truncated for $\rho<1$.
Denoting $\Lambda=\log \rho/ \log n$, $\tilde{\mu}=\mu_{\log \rho|n}/\log n$ and $\tilde{\sigma}^2=\sigma_{\log \rho|n}^2/\log n$, this expression becomes
\begin{equation}
\mu_{\rho^{-q} | n} \approx \int_1^\infty e^{H(\Lambda)\log n } \text{d}\rho,
\end{equation}
with $H(\Lambda)=-(\Lambda-\tilde{\mu})^2/(2\tilde{\sigma}^2)-q\Lambda$. For large $n$, the value of this integral tends to $e^{H(\Lambda^*)\log n}$ where the $\Lambda^*$ is the value where $H$ takes a maximum, that is either at  $\Lambda^*=\tilde{\mu}-q\tilde{\sigma}^2$ if $\tilde{\mu}-q\tilde{\sigma}^2>0$, or $\Lambda^*=0$ otherwise. Thus,
\begin{eqnarray}
\log \mu_{\rho^{-q} | n} & \approx & - ( {\gamma}_{q,\rho^{-1}|n} \log n +{\omega}_{q,\rho^{-1}|n}) \\ 
& \text{ with } &
\left\lbrace
\begin{array}{ll}
{\gamma}_{q,\rho^{-1}|n}= q\tilde{\mu} - q^2 \tilde{\sigma}^2 / 2 & \text{ if }  \tilde{\mu} > q \tilde{\sigma}^2, \\
{\gamma}_{q,\rho^{-1}| n}= \tilde{\mu}^2/(2\tilde{\sigma}^2) & \text{ if }  \tilde{\mu} \leq q \tilde{\sigma}^2,
\end{array}
\right. \nonumber
\end{eqnarray}
and ${\omega}_{q,\rho^{-1}|n}=q \log (\mathcal{A}/(s_a \ell_0 ))$ a constant.  In particular, we are interested in the exponent $q=2$ which is useful to describe fluctuations around the mean. We have 
\begin{equation}
\tilde{\gamma} \equiv {\gamma}_{2,\rho^{-1}|n} = \left\lbrace \begin{array}{c}
2 \tilde\mu_\lambda- 2 \tilde\sigma^2_\lambda  \text{ if }  \tilde{\mu} > 2 \tilde{\sigma}^2 \\
\tilde\mu^2/(2\tilde\sigma^2)   \text{ if }  \tilde{\mu} \leq 2 \tilde{\sigma}^2
\end{array} \right. .
\label{eq:tildegamma}
\end{equation}
The predicted dependence of $\tilde{\gamma}$ upon $D_1$ is reproduced in Fig.~\ref{fig:scaling_theta}. $\tilde{\gamma}$ is bounded between 2 (for $D_1 \to 2$) and 1 for $D_1\approx 1.5$. %For $D_1<2$, $1<\tilde\gamma<2$. For $D_1> 1.75$, $\tilde\gamma$ follows the typical moments of a non-truncated log-normal distribution. Below this value, the probability of weak stretching rates $P_{\rho|n}(\rho=1)$ entirely determines the scaling of statistical moments.  
The prediction agrees reasonably well with numerical simulations of the baker map and sine flow (Fig.~\ref{fig:scaling_theta}). The small discrepancies can be attributed to deviations from log-normally distributed elongation in bundles. We also verify numerically that $\tilde\omega \equiv \omega_{2,\rho^{-1}|n}$ is independent of $D_1$ (Fig.~\ref{fig:scaling_theta}b). 
In Fig.~\ref{fig:gamma2_omega2}, we verified that $\tilde\gamma$ is independent of the aggregation scale $s_a$.  In contrast, 
\begin{equation}
\tilde\omega \approx 2 \log \left(\mathcal{A}/(l_0 s_a) \right), 
\label{eq:omega2}
\end{equation}
 is a function of the aggregation scale only, independent of time and fractal dimension (Fig.~\ref{fig:scaling_theta}.b).
Having determined both the elongation statistics inside aggregates of size $n$ and the spatial distribution of $n$, we will deduce in the following section the statistics of aggregated scalar levels $c$.

\section{Aggregated scalar concentrations}\label{sec:unconditional}
In the preceding section, we have shown that the moments of lamella compression inside a bundle of size $n$ follow:
\begin{eqnarray}
\mu_{\rho^{-1} | n} &=& \frac{s_a \ell_0}{\mathcal{A}} n^{-1} \label{eq:meanrho-1}\\
\mu_{\rho^{-2} | n} &=& \frac{(s_a \ell_0)^2}{\mathcal{A}^2} n^{-\tilde{\gamma}}
\label{eq:statisticsbundles}
\end{eqnarray}
with $\tilde{\gamma}\equiv\gamma_{2,\rho^{-1}|n}$ a flow-dependent exponent varying with the fractal dimension $D_1$ and taking value between 1 and 2 (Fig.~\ref{fig:scaling_theta}a). The variance of lamellar concentrations inside bundles thus follows:
\begin{eqnarray}
\sigma^2_{\rho^{-1}|n} &=& \frac{(s_a \ell_0)^2}{\mathcal{A}^2} \left(n^{-\tilde{\gamma}}- n^{-2} \right).
\label{eq:variance}
\end{eqnarray}

We first assume that the aggregated scalar concentration of a bundle $c|n$ can be obtained by a sum of independent and identically distributed random variables (Eq.~\eqref{eq:sum}), whose statistics have been described previously. 
In doing so, we make two hypothesis. First, we assume that the elongation statistics of lamella inside a bundle of size $n$ or among all bundles size $n$ are comparable, which is true for large $n$ (see Appendix ~\ref{sec:sampling}). Second, we assume that variability in bundle aggregated concentrations arises from \textit{independent} realizations of the sum. In other words, we assume that the chaotic flow is sufficiently random to shuffle lamellar elongations between each bundles. We will show later that independence is not ensured in the deterministic Baker map.

With these hypothesis, the mean aggregated concentration is
\begin{eqnarray}
	\mu_{c|n} =  n  \cdot \frac{\sqrt{\pi}  \theta_0 s_0}{s_a} \mu_{\rho^{-1} |n}  = \frac{\sqrt{\pi}  \theta_0 \ell_0 s_0}{\mathcal{A}}, % \\
	%sigma^2_{c|n} &=& n \cdot \left( \frac{\theta_0 s_0}{s_a} \right)^2 \sigma^2_{\rho^{-1}|n}  = \frac{(\theta_0 \ell_0 s_0)^2}{A^2} \left(n^{1-{\gamma}_{2,\rho^{-1}|n}}- n^{-1} \right), \label{eq:independent}
\end{eqnarray}
which is also the mean concentration $\mu_c$. The variance of aggregated concentrations over bundles of similar size is
\begin{equation}
\sigma^2_{c|n} \approx n  \left(\frac{\sqrt{\pi} \theta_0 s_0}{s_a}\right)^2  \sigma^2_{\rho^{-1}|n} =\frac{(\sqrt{\pi}  \theta_0 \ell_0 s_0)^2}{\mathcal{A}^2} \left(n^{1-\tilde{\gamma}} -  n^{-1} \right).
\label{eq:xi}
\end{equation}
When $n$ large, this expression further simplifies to
\begin{equation}
\sigma^2_{c|n} \sim n^{1-\tilde{\gamma}},
\label{eq:independent}
\end{equation}
with $\tilde{\gamma} \in [1,2]$ given by Eq.~\ref{eq:tildegamma}. 

\begin{figure}
\centering \includegraphics[width=6cm]{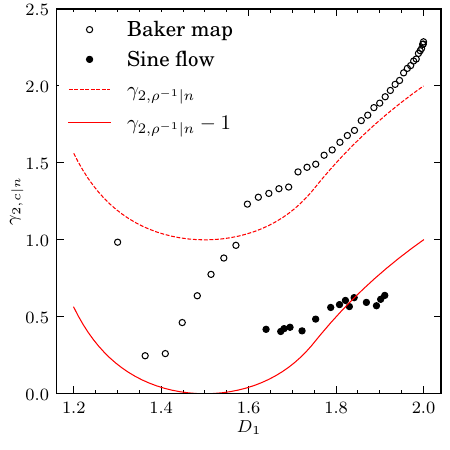}
\caption{Scaling exponent $\xi$ (Eq.~\eqref{eq:c|n}) of the variance of bundle concentrations knowing $n$ estimated from simulations (dots) and theoretical predictions with the independent realisation hypothesis for the sine flow (dashed lines, $\xi=\tilde\gamma - 1 $, Eq.~\eqref{eq:independent}) and baker map (continuous lines, $\xi =\tilde\gamma $, Eq.~\eqref{eq:xi_dependent}). }  \label{fig:independent}
\end{figure}

The scaling obtained with the independent realization hypothesis compares well with numerical simulations of aggregation in random sine flows of various heterogeneity (Fig.~\ref{fig:independent}). However, it largely underestimates the scaling observed in deterministic baker map. Indeed, the simplicity and regularity of the deterministic baker map makes bundles of similar size not statistically independent. While bundle concentration still results from the addition of variable lamellar concentrations, independent realisations of the sum are not achieved due to the deterministic nature of the map, the same combinations being repeated identically in most bundles, as in the case of a unique realization. In that case, the variance of the sum is the variance of the random variable and Eq.~\eqref{eq:xi} transforms into
\begin{equation}
\sigma^2_{c|n} \sim  \sigma^2_{\rho^{-1},n} \sim n^{-\tilde{\gamma}}.
\label{eq:xi_dependent}
\end{equation}
This scaling indeed fits more accurately the deterministic baker map simulations (Fig.~\ref{fig:independent}). 

To summarise, the addition of lamellar concentration levels in a bundle yields a concentration whose deviation from the mean decays algebraically with the number of lamella in the bundle 
\begin{eqnarray}
\sigma^2_{c | n} &\approx& \frac{(\theta_0 \ell_0 s_0)^2}{\mathcal{A}^2}  n^{-\xi}  ,
\label{eq:c|n}
\end{eqnarray}
with $\xi=\tilde{\gamma}$ for purely deterministic flows (baker map) and $\xi=\tilde{\gamma}-1$ for random flows (sine flow). We call $\xi$ the \textit{correlation} exponent, which can take values between 0 and 2 depending on the flow heterogeneity and randomness.

\subsection{Distribution of $c$}
We now focus on deriving the expression of the pdf of aggregated scalar concentration $P_c$ over the flow domain.  The later can be obtained from the conditional pdf of bundle aggregated concentrations knowing their size $P_{c|n}$ and the pdf of bundles size $P_n$ by the weighted summation
\begin{equation}
P_c(c) = \int_n \text{d}n P_{c|n}(c) P_n(n).
\label{eq:prediction_pdf}
\end{equation} 
$P_{n}$ was found to be well approximated by a Gamma distribution, with moments given by the fractal characteristics of advected material lines (Eq.~\eqref{eq:parameters_n}). In turn, we have determined the scaling of the first two moments of bundle aggregated concentration $c|n$ (Eq.~\eqref{eq:c|n}), without specifying the precise shape of the pdf. A natural choice for $P_{c|n}$ is the log-normal distribution, since the sum of log-normally distributed random variables are known~\citep{schwartz1982distribution} to be well approximated by log-normal distributions. Adopting the log-normal shape, the parameters of the distribution are
\begin{eqnarray}
\mu_{\log c|n} &=& \log\mu_{c|n} - \log(\sigma^2_{c|n}/\mu_{c|n}^2+1)/2 \\
\sigma^2_{\log c|n} &=& \log(\sigma^2_{c|n}/\mu_{c|n}^2+1).	
\end{eqnarray}

\begin{figure}
\includegraphics[width=0.49\linewidth]{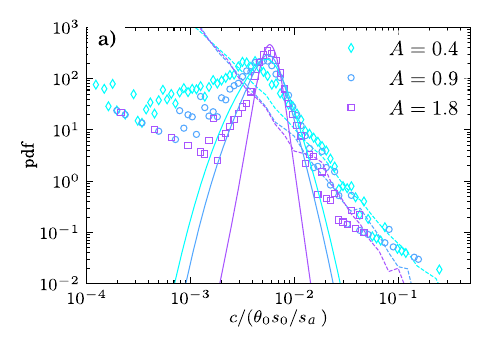}
\includegraphics[width=0.51\linewidth]{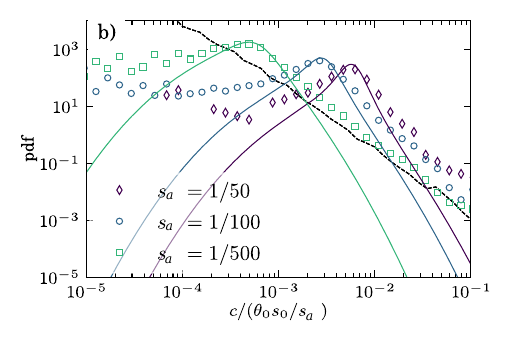}
\caption{Distributions of aggregated scalar concentrations in the sine flow depending on a) the sine wave amplitude $A$ ($s_a=1/50$) and b) the aggregation scale $s_a$ ($A=0.9$). Dots stand for numerical simulations, continuous lines are the aggregation model (Eq.~ \eqref{eq:prediction_pdf}), and the dashed lines are the isolated strip prediction (Eq.~\eqref{eq:isolatedstripdistribution}). Simulations are all taken at the time when the total filament length reaches $L=10^7 \ell_0$.}\label{fig:pdf_c}
\end{figure}

In Fig.~\ref{fig:pdf_c}, we plot the simulated distribution of aggregated concentration levels compared to the prediction Eq.~\eqref{eq:prediction_pdf} for the baker map and sine flow. The agreement is fair in the region near $\mu_c$, but deviates for large $c$. Indeed, this corresponds to lamellae with weak aggregation for which $n\approx 1$. In this region, the solitary strip pdf $P_{\rho^{-1}}$ describes well the tail of $P_c$ because such high concentration excursions are essentially supported by isolated lamellae, while the correlated aggregation model assumes $n\gg1$. The presence of these weakly aggregated, high concentration levels is particularly evident at high Péclet number (Fig.~\ref{fig:pdf_c} (b)). 
The scalar concentration pdf is thus the combination of an aggregated core around the mean following Eq.~\eqref{eq:prediction_pdf} and tails following the isolated strip concentration pdf.
% \textcolor{blue}{quand est il des statistiques aux basses concentrations ? Est ce qu'elles ne seraient pas mieux décrit par une gamma comme elles sont caractérisées par des lois de puissance ?}

\begin{figure}
\centering\includegraphics[width=8cm]{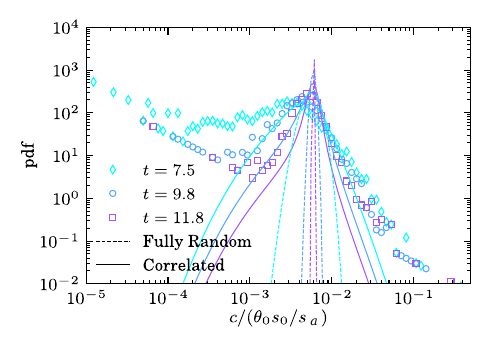}
\caption{$P_c$ in sine flows at several times ($A=0.8, s_a=1/50$) (dots) compared with the random aggregation (dashed lines) and correlated aggregation (continuous lines, Eq.~\eqref{eq:prediction_pdf}) models.}\label{fig:pdftime}
\end{figure}

In Figs.~\ref{fig:sinePDF}c and \ref{fig:pdftime}, we compare the correlated aggregation model with the random aggregation model where $\mu_n \sim L(t)$ (Eq.~\eqref{eq:n_estimate}).  The random aggregation assumption yields gamma pdfs that are narrowing much faster than the simulated pdfs in the sine flow. In contrast, the correlated model better captures the pdf. 
From the pdf of aggregated scalar concentration, we now derive its moments.  They are directly related to the pdf of $n$, since
\begin{eqnarray}
\mu_c &=& \int \text{d}c \sum_n  c P(c|n) P(n) =  \sum_n  \mu_{c|n}  P(n) =   \frac{\theta_0 \ell_0 s_0}{\mathcal{A}}
\label{eq:c}
\end{eqnarray}
and
\begin{eqnarray}
\mu_{c^2} &=& \int_c \text{d}c \int_n  \text{d}n \, c^2 P(c|n) P(n)   = \int_n\,   \text{d}n  \mu_{c^2|n}  P(n)  =  \frac{(\theta_0 \ell_0 s_0)^2}{\mathcal{A}^2} ( \mu_{n^{-\xi}} +1).
\label{eq:c2}
\end{eqnarray}
%\red{Initial variance of $c$ at $n=1$? How to introduce transient with small $n$?}
Thus, the scalar variance is
\begin{eqnarray}
\sigma^2_c &=&  \frac{(\theta_0 \ell_0 s_0)^2}{\mathcal{A}^2}  \mu_{n^{-\xi}}.
\label{eq:sigmac2}
\end{eqnarray}

\begin{figure}
	\centering \includegraphics[width=8cm]{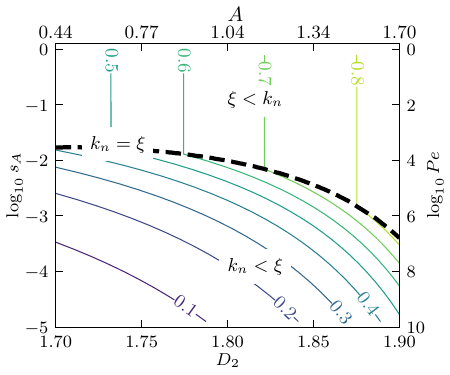}
	\caption{Evolution of the exponent $\text{min}(k_n,\xi)$ (contours) in the sine flow for various Péclet (aggregation scale $s_A$) and sine amplitude $A$ (fractal dimension $D_2$). $k_n$ is obtained with Eq.~\eqref{eq:kn} and $\xi=\tilde{\gamma}-1$ with Eq.~\eqref{eq:tildegamma}}
	\label{fig:exponent}
\end{figure}

Note that, because we chose a gamma distribution for $n$ with parameters $\theta_n$ and $k_n$ defined in Eqs.~\eqref{eq:parameters_n},  $\mu_{n^{-\xi}}$ is not defined for all $D_2$.
Indeed, the gamma distribution admits a power law tail at small $n$ with exponent $k_n-1$, which affects the value of negative moments: when $k_n(D_2)<\xi(D_2)$, $\mu_n^{-\xi}$ does not exist. 
However, because fully chaotic flows are space-filling $(D_0=2)$, we must have $n\geq 1$ at late time.  Thus, we restrict the pdf of $n$ to $n\geq1$ and get 
\begin{equation}
 \mu_{n^{-\xi}} \sim \left(\theta_n\right)^{-\text{min}(k_n,\xi)}
\label{eq:n_xi}
\end{equation}
An intuitive understanding of this equation can be formulated as follows. If the spatial heterogeneity of $n$ is moderate ($\xi<k_n$), the average of $n^{-\xi}$ is affected by all values of $n$ in the distribution. In contrast, if the heterogeneity is stronger ($\xi>k_n$), the probability of having low aggregation regions ($n\approx 1$) is high and controls the value of $ \mu_{n^{-\xi}}$. In that case, the average does not scale anymore with $\xi$, but rather with the parameter $k_n$, explaining the minimum in the exponent. The value of the exponent depending on Péclet number and fractal dimension is plotted in Figure~\ref{fig:exponent} for the sine flow. In such random flow, the exponent takes values between 0 and 1. For high Péclet numbers and low fractal dimension, $\mu_{n^{-\xi}}$ is governed by weakly aggregated regions and $\min(k_n,\xi)=k_n$. In contrast, for low Péclet number and large fractal dimension, the whole distribution of $n$ plays a role in the determination of $\mu_{n^{-\xi}}$ and $\min(k_n,\xi)=\xi$.

Combining Eq.~\eqref{eq:n_xi} and Eq.~\eqref{eq:parameters_n} provides the asymptotic scalar variance decay  as a function of the domain area $\mathcal{A}$, the elongation of material lines $L(t)$, and the aggregation scale $s_a=\sqrt{2\pi}s_B$:
\begin{equation}
\sigma^2_c(t) =  \left( \frac{L(t)s_a (\sqrt{\mathcal{A}} s_a^{D_2-2}-1)}{\mathcal{A}}\right) ^{-\text{min}(k_n,\xi)} \sim L(t)^{-\text{min}(k_n,\xi)}.
\label{eq:sigma_c_correlated}
\end{equation}
Thus, in a correlated aggregation scenario, the decay exponent of scalar variance is found to be a fraction of the growth exponent of material lines; respectively $\log 2$ and $\mu_\lambda+\sigma^2_\lambda/2$ in the baker map and sine flow. %Note that in sine flow, in general in sine flow, $k_n \geq \xi$ such that the scalar variance decay exponent is $\gamma_{c,2}=(\mu_\lambda+\sigma_\lambda^2/2) \xi$.

Eq.~\eqref{eq:sigma_c_correlated} suggests that the existence of correlations in the aggregation process may be envisioned as a retardation of the purely random aggregation scenario (Eq.~\eqref{eq:random}). From this perspective, the effective number of \textit{random} aggregation events $L(t)^{\text{min}(k_n,\xi)}$ is smaller than the total number of aggregation $L(t)$, because the correlation exponent $\text{min}(k_n,\xi)$ is generally smaller than one (see Fig.~\ref{fig:exponent}).

In contrast, the random aggregation model (Eq.~\ref{eq:random}) clearly overestimates the variance decay rate in the sine flow. This is explained by the correlated nature of aggregation which is less efficient at homogenising concentration levels than a completely random addition. In other words, small concentration levels have a higher probability of coalescing with other small concentrations than with high concentrations, retarding the homogeneisation of the mixture. Indeed, the correlated aggregation model proposed herein (Eq.~\eqref{eq:sigma_c_correlated}) shows decay rates closer to the numerical observations. Solitary lamella model still outperforms the correlated aggregation framework in predicting scalar decay rate, because the latter is dominated by weakly aggregated regions of the flow. These regions are poorly described by statistics obtained in the large-$n$ limit (e.g. Eq.~\eqref{eq:xi}). 

%The asymptotic scalar decay rate is thus driven almost entirely by the evolution of the stretching statistics and solitary strip concentration levels, the aggregation being too correlated and inefficient to accelerate mixing. %This is also why the fully correlated model, which is entirely described by the stretching pdf of solitary lamellae (Eq.~\eqref{eq:fullycorrelated}), accurately captures the variance decay rate (Fig.~\ref{fig:gamma2}).

\begin{figure}
\includegraphics[width=0.49\linewidth]{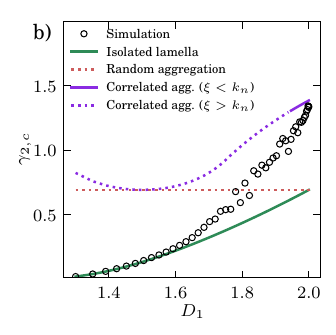}
\includegraphics[width=0.51\linewidth]{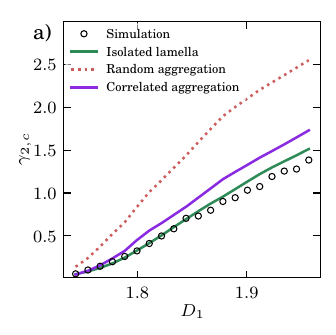}
\caption{Decay exponent $\gamma_2$ of the variance of aggregated scalar levels with time, as a function of fractal dimension $D_1$ for the a) random sine flow and b) baker map. Dots stand for numerical simulations and lines from theoretical predictions for isolated lamellae (Eq.~\eqref{eq:variance_sol} and exponents of $\mu_{\rho^{-1}}$ in Tables~\ref{tab:table1} and \ref{tab:table2}), random aggregation (Eq.~\eqref{eq:random} and exponents of $1/\mu_{\rho}$ in Tables~\ref{tab:table1} and \ref{tab:table2}) and correlated aggregation (Eq.~\eqref{eq:sigmac2}). }\label{fig:gamma2}
\end{figure}

A different picture is observed for scalar decay rates obtained in the baker map (Fig.~\ref{fig:gamma2}b), because of the deterministic nature of the map. The solitary lamella model is accurate for fractal dimensions $D_1<1.6$, where stretching heterogeneity is important and weakly aggregated regions of the flow dominate scalar fluctuations.
For $D_1>1.6$, the isolated lamella model underpredicts scalar decay due to aggregation effects as explained in the following. The random aggregation scenario predicts an invariant scalar decay rate ($\gamma_{2,c} = \log 2$) equal to the growth rate of material lines. It thus overpredicts the observed decay rates for $D_1<1.8$ and underpredicts it at larger values of $D_1$. 
Interestingly, when the flow tends to the uniform case $a\to 0.5$ ($D_1\to 2$), simulations yields scalar decay rates of $2\log2$, larger than the decay rate of isolated lamella. This acceleration of mixing by aggregation is a pure consequence of the determinism nature of the map, and is well captured by the correlated aggregation scenario (Eq.~\eqref{eq:xi_dependent}).  However, for all baker maps with $D_1<1.9$ ($a<0.2$), $k_n<\xi$ and scalar fluctuations are mainly governed by the regions where $n\sim 1$ for which the correlated aggregation theory is not expected to be accurate. This explains the limited range of validity of the correlated aggregation theory for describing scalar decay rates in the deterministic baker map.

Finally, note that the simulations do not show the super-exponential decay of scalar fluctuations classically observed for the uniform stretching rate at $a= 0.5$. In fact, the reconstruction of the scalar field by a summation of lamellar maximum concentrations on a fixed grid (Eq.~\eqref{eq:aggconcentration}) impedes the apparition of the super-exponential mode. As $a\to 0.5$, all lamella are subjected to similar stretching rates around $\log 2$, thus yielding a scalar variance decaying as $2\log2$.

\section{Conclusions}
Scalar mixing by incompressible smooth two-dimensional chaotic flows leads to the emergence of two coupled phenomena which are the elongation of interfaces by stretching, and their aggregation and coalescence at the Batchelor scale by compression. In turbulent scalar mixing, aggregation was well captured by random addition rules~\citep{DuplatVillermaux08}. Here, we show that for smooth chaotic flows in the Batchelor range, aggregation is highly correlated. This correlated aggregation process significantly reduces the flow mixing efficiency compared to a random hypothesis, maintaining it close to the mixing efficiency obtained for solitary lamellae and explaining the observed monotonic exponential decay of scalar variance before and after coalescence time~\citep{fereday2002scalar}.
%by chaotic flows results from the coupling of exponential fluid deformations and molecular diffusion. At large times, elongation in a finite domain forces elongated scalar filaments to wrap around each other and coalesce at the Batchelor scale. The linearity of the advection-diffusion equation offers the possibility to describe the evolution of the scalar field as the summation of individually stretched lamellae aggregated into bundles~\citep{villermaux2003mixing}. 

%In this paper, we uncover the statistical property of scalar aggregation in two-dimensional periodic chaotic flows. In particular, we show that the aggregation process is strongly heterogeneous and correlated to local stretching rates. This confirms previous observations~\citep{HeymanPRL2020} that strongly elongated lamellae are also the most aggregated, and vice versa. 
Using two-dimensional chaotic flows as a reference, we measured the aggregation rate of exponentially stretched material lines across a broad range of chaotic flow regimes. We showed that the most elongated lamellae are also the most aggregated ones, due to the fact that larger compression rates attract larger flow regions. The link between elongation and compression, induced by incompressibility, hence generates a direct correlation between elongation and aggregation. The heterogeneity in stretching rates therefore controls the heterogeneity of the number of lamellae in bundles.

We showed that the statistics of aggregated lamella numbers can be exactly predicted from the fractal dimensions of the elongated material line. We then derived a general theoretical framework that captures the effect of correlated aggregation, where lamellae of similar stretching aggregate preferentially, and predict the pdfs of aggregated scalar levels.
%We also demonstrate that the stretching statistics of lamellae inside bundles of similar size are tight to the information dimension. 
%Such characterization of local stretching statistics allows for the prediction of pdf of aggregated scalar levels, for a large range of flow heterogeneity.
We find that correlated aggregation can be uniquely characterised by single correlation exponent $\xi \in [0,1]$, which provides a measure of the \textit{effective} number of aggregation events, compared to the \textit{total} number. In that sense, correlated aggregation delays the route to uniformity compared to a fully random hypothesis, although it does not alter the fundamental nature of the aggregation process~\citep{villermaux2003mixing}.

Our results apply for two-dimensional fully chaotic flows in the Batchelor regime, that is, for smooth velocity fields below the integral scale. These flow fields are representative of a large class of flows, including notably porous media flows \citep{Heyman2020, Souzy:2020aa}, geophysical flows, and turbulent flows at high Schmidt number. 
%It is probable that different aggregation rules arise in rough flows above the integral velocity scale, which remain well captured by random aggregation rules~\citep{DuplatVillermaux08}. A remaining open question is thus to uncover the potential mechanisms leading to a loss of correlations from small diffusive scales to large dispersive scales.
We have also considered small periodic flow domains where no large scale scalar gradients can appear. When scalar length scales can develop beyond the integral velocity scale in smooth flows, mixing is controlled \textit{globally} by the slowest dispersing modes~\citep{haynes2005controls,tsang2005exponential}. The aggregation rules determined herein are not supposed to change above the velocity scale. Thus, it is in principle possible to obtain the statistics of larger scale scalar fields by the summation of individual lamella. To do so, one may consider the mean density of aggregation $\mu_n(t)$ (Eq.~\eqref{eq:mun}) to vary spatially through macrodispersion, while microscale fluctuations would remain governed by Eq.~\eqref{eq:sigman}. Note that similar ideas were used to predict the evolution of a dispersing scalar plume in a two-dimensional porous media~\citep{LeBorgneJFM2015}.

It should also be possible to extend the correlated aggregation theory to three-dimensional flows that produce two-dimensional sheets ~\citep{martinez2018diffusive,ngan2011scalar,meunier2022diffuselet}, that aggregate at late time due to folding. These flows have a unique negative Lyapunov exponent, the mean compression rate normal to the sheets, and thus also simplify to the one-dimensional Lagrangian stretching framework on which aggregation models are framed.

%Discussion on the consequence of taking a fix coarsening scale : In fact, the coarse graining size of the resulting concentration field may vary from one position to another, since some regions are more stretched than others. This said, the average fluctuations lengthscale is fixed at the Batchelor scale and fluctuations around this value are small. Thus, $s_a$ constitute a natural lengthscale to determine the local statistics of elongation in bundles.

 %The random \textit{anzatz} for constructing the scalar concentration pdf by $n$-convolution product of the single lamellar pdf is thus not directly applicable for these flows, unless an arbitrary $n_\text{eff} = \langle c\rangle ^2/ \langle c^2\rangle \ll \langle n(t) \rangle$ is chosen.
%In contrast to the random aggregation scenario, we find that aggregation is not the main responsible for the homogenization of a mixture by chaotic mixing, but rather the decay of low stretching events. In particular, the perfectly correlated scenario \eqref{p(thetaB)} induces a net decay of scalar fluctuations by the sole convergence to the mean $\langle c \rangle$, while conserving the shape of the single lamella lognormal pdf~\eqref{eq:lognorm} at large concentration. Fast (super-exponential) homogeneization of scalar fluctuations do occur closer to the the mean concentrations, although they do not dominate the evolution of the scalar variance. Our results gives a physical interpretation of the validity of Lagrangian stretching theoretical framework after coalescence time.

%\section{References}
%\footnotesize
%\setlength{\bibsep}{0.0pt}
%
\bibliographystyle{jfm}
%\bibliography{biblio.bib}

%\iffalse

%\fi

\backsection[Acknowledgements]{JH, TLB and PD acknowledge the Europe Research Council under the grant 101042466 (CHORUS). TLB and EV acknowledge the H2020 MSCA ITN program, under the grant 956457 (COPERMIX).  Three anonymous reviewers are thanked for contributions to the manuscript.}

\backsection[Funding]{Funded by the European Union (ERC, CHORUS, 101042466)}

\backsection[Declaration of interests]{The authors report no conflict of interest.}

\backsection[Data availability statement]{The data and codes to reproduce the findings of this study are openly available in Github at https://github.com/jorishey1234/aggregation or upon reasonable request to joris.heyman@univ-rennes.fr}

\backsection[Author ORCIDs]{J. Heyman, https://orcid.org/0000-0002-0327-7924; 
T. Le Borgne, https://orcid.org/0000-0001-9266-9139;
 E. Villermaux, https://orcid.org/0000-0001-5130-4862;  P. Davy, https://orcid.org/0000-0002-6648-0145}

\appendix

\section{Stretching statistics and averaging}\label{sec:SM_average}
%\begin{figure}
%\centering \includegraphics[width=8cm]{figure18.png}
%\caption{Different sampling procedures a. Uniformly over the initial filament b. uniformly over the final elongated filament} \label{fig:sampling}
%\end{figure}

In random chaotic flows, varying stretching rates are experienced by fluid elements~\citep{Lester2013}. Because of the multiplicative nature of stretching, the log-elongation of material elements $\log \rho$ is well approximated in ergodic chaotic flows by a sum of iid random variables, that converges towards the normal distribution with mean $\mu_\lambda t$ and variance $\sigma^2_\lambda t$~\citep{meunier2022diffuselet}.

Thus, the elongation $\rho$ of material elements may be expected to follow the log-normal distribution
\begin{equation}
P_{\rho,0}(\rho) \approx \frac{1}{\rho\sqrt{2\pi \sigma^2_\lambda t}} \exp\left( - \frac{(\log \rho - \mu_\lambda t)^2}{2\sigma^2_\lambda t} \right)\label{eq:lognorm}.
\end{equation}

Non-asymptotic stretching statistics can differ substantially from this limiting behavior. For instance, the baker map has a binomial distribution of elongations 
\begin{equation}
P_{\rho,0}\left(\rho=(1-a)^k a^{t-k}\right) = \left(\begin{array}{c}
t \\
k
\end{array} \right) (1-a)^k a^{t-k}, \text{ for } k=0 \dots t,
\end{equation}
which tends to a lognormal distribution with $\mu/t=-a\log(a)-(1-a)\log(1-a)$ and $\sigma_\lambda^2/t=a (1-a) (\log(1-a) - \log(a))^2$.

Note also that, because of the multiplicative nature of elongation~\citep{redner1990random}, the pdf of $\rho$ (and $\rho^{-1}$) is highly sensitive to the tails of $P_{\log \rho}$, which may converge only slowly to the Gaussian prediction depending on the flow heterogeneity. However, for the range of sine flow amplitudes used in this study ($A\in [0.4,1.8]$), such limitations are relatively weak.

Statistics of lamellar concentrations in the material line can then be obtained by suitable ensemble averaging (denoted by angle brackets) over this distribution. Depending on how sampling is performed through the material line, different moments are obtained. Uniform sampling on the \textit{initial} filament prior elongation leads to the distribution $P_{\rho}(\rho)$ (Eq.~\eqref{eq:lognorm}) with main moments summarized in Table~\ref{tab:table1}. In contrast, uniform sampling on a the \textit{final} elongated material line of length $L$, denoted by $\mu_{\bullet,L}$ leads to the weighted pdf  $P_{\rho,{L}}\sim \rho P_{\rho}$. $P_{\rho,{L}}$ is thus also lognormal, but with different mean, $(\mu_\lambda+\sigma^2_\lambda) t$. Uniform sampling on the final material line gives a stronger weights to highly elongated part of the material line than uniform sampling on the initial filaments. Moments of $\log \rho$, $\rho$ and $\rho^{-1}$ are summarized in Tables~\ref{tab:table1} for sine flow and \ref{tab:table2} for the baker map for initial and final sampling. 
Note that we impose $\rho\geq 1$ since the one-dimensional lamellar framework is valid only when lamellae elongates in the $y$ direction. This lower bound creates a particular scaling of moments of $\rho^{-1}$ when $\sigma_\lambda^2$ is larger than $\mu$, that is when weak stretching rates dominate the ensemble average.

\begin{table}
\def~{\hphantom{0}}
\begin{center}
\begin{tabular}{p{1cm}p{3cm}p{3cm}p{5cm}}

~ 								& $\log \rho$ 	& $\rho$ 		& $\rho^{-1} $   \\
\hline
$\mu_{\bullet}$  		&$\mu_\lambda t$		&$e^{(\mu_\lambda+ \sigma_\lambda^2/2)t}$		& $\left\lbrace \begin{matrix} e^{-(\mu-\sigma_\lambda^2/2)t} &\text{ if } \mu_\lambda\geq \sigma^2_\lambda \\ e^{- \mu^2/(2\sigma_\lambda^2)t} &\text{ if } \mu_\lambda\leq \sigma_\lambda^2 \end{matrix} \right.$	\\
\hline
$\mu_{\bullet^2}$  	&$\sigma^2_\lambda t$		&$e^{(2\mu_\lambda+ 2\sigma^2_\lambda)t}$		&  $\left\lbrace \begin{matrix} e^{-2(\mu-\sigma^2_\lambda)t} &\text{ if } \mu_\lambda\geq 2\sigma^2_\lambda \\ e^{- \mu^2/(2\sigma_\lambda^2)t} &\text{ if } \mu_\lambda\leq 2\sigma^2_\lambda \end{matrix} \right.$	\\
\hline
$\mu_{\bullet,L}$  	& $(\mu+\sigma^2_\lambda) t$ &$e^{(\mu_\lambda+ 3\sigma^2_\lambda/2)t}$		&$e^{(-\mu_\lambda- \sigma^2_\lambda/2)t}$	\\
\hline
$\mu_{\bullet^2,L}$ 	& $\sigma^2_\lambda t$			&$e^{(2\mu_\lambda+ 4\sigma^2_\lambda)t}$		& $\left\lbrace \begin{matrix} e^{-2\mu_\lambda t} &\text{ if } \mu_\lambda\geq \sigma^2_\lambda \\ e^{- (\mu+\sigma^2_\lambda)^2/(2\sigma^2_\lambda)t} &\text{ if } \mu_\lambda\leq \sigma^2_\lambda \end{matrix} \right.$	 	\\

\end{tabular}
\caption{Moments of log-normally distributed stretching sampled over infinitesimal fluid elements ($\langle  \bullet \rangle_0$), material line ($\langle  \bullet \rangle_L$) . }\label{tab:table1}
\end{center}
\end{table}

\begin{table}
\def~{\hphantom{0}}
\begin{center}
\begin{tabular}{p{1cm}p{5cm}p{3cm}p{3cm}}
~ 								& $\log \rho$ 	& $\rho$ 		& $\rho^{-1}$   \\
\hline
$\mu_{\bullet}$  		&$ (- a\log(a) - (1-a)\log(1-a))t$		&$2^t$		& $(1-2a +2a^2)^t$	\\
\hline
$\mu_{\bullet^2}$  	&$ a(1-a) (\log(1-a) - \log(a))^2 t$		&$-$		&  $(1-3a +3a^2)^t$	\\

%$\langle \bullet \rangle_L$ 	& $(\mu+\sigma^2) t$ &$e^{(\mu_\lambda+ 3\sigma^2/2)t}$		&$e^{(-\mu_\lambda- \sigma^2/2)t}$	\\
%$\langle \bullet^2 \rangle_L$ 	& $\sigma^2_\lambda t$			&$e^{(2\mu_\lambda+ 4\sigma^2)t}$		& $\left\lbrace \begin{matrix} e^{-2\mu_\lambda t} &\text{ if } \mu_\lambda\geq \sigma^2_\lambda \\ e^{- (\mu+\sigma^2)^2/(2\sigma^2)t} &\text{ if } \mu_\lambda\leq \sigma^2_\lambda \end{matrix} \right.$	 	\\	
%\hline
\end{tabular}
\caption{Moments of binomial distributed stretching sampled over infinitesimal fluid elements ($\langle  \bullet \rangle_0$), material line ($\langle  \bullet \rangle_L$).}\label{tab:table2}
\end{center}
\end{table}

\section{Fractal dimensions in the baker map}~\label{app:baker}

The fractal dimension of order $q$ of the measure $p$ is obtained with~\citep{grassberger1983generalized}:
\begin{equation}
	D_q -1 = \lim_{\epsilon \to 0} \frac{1}{q-1} \frac{ \log I_{q}(\epsilon)}{\log \epsilon}, \quad
	I_q(\epsilon) \equiv \sum_k^{N=\mathcal{L}/\epsilon} p_k^q,
\end{equation}
where the subtraction of 1 on the left hand side accounts for the counting of one-dimensional structures (lamellae) in a two-dimensional domain.  This definition implies the following  spatial scaling  of the integral of the measure:
\begin{equation}
	I_q(\epsilon) \sim \epsilon^{(q-1)(D_q-1)}.\label{eq:scalingIq2}
\end{equation} 
In the baker map, te integral of the measure can then be computed by summing its value on the two replicates created by the map,
\begin{equation}
	I_q(\epsilon) = I_{q,a}(\epsilon) + I_{q,1-a}(\epsilon).
	\label{eq:Iq}
\end{equation}
We observe that
\begin{equation}
	I_{q,a}(a\epsilon)=I_{q,1-a}((1-a)\epsilon)=\sum^{N=1/\epsilon}_k \left( \frac{p_k}{2} \right)^q,
\end{equation}
where the factor $1/2$ comes from the normalisation of the measure due to the doubling of $n$. Thus
\begin{equation}
	I_{q,a}(\epsilon)=I_{q}(\epsilon/a) 2^{-q} \mbox{ and } I_{q,1-a}(\epsilon)=I_{q}(\epsilon/(1-a)) 2^{-q}.
\end{equation}
Replacing the last expression in \eqref{eq:Iq} yields
\begin{equation}
	I_q(\epsilon) = 2^{-q} \epsilon^{(q-1)D_q} \left( {a}^{-(q-1)(D_q-1)} +(1-a)^{-(q-1)(D_q-1)}\right)
\end{equation}
Using the scaling $I_q(\epsilon) \sim  \epsilon^{(q-1)D_q}$, thus provide a transcendental equation for $D_q$ independently of $\epsilon$:
\begin{equation}
	2^q= \left( {a}^{-(q-1)(D_q-1)} +(1-a)^{-(q-1)(D_q-1)}\right),
	\label{eq:Fractalbakermap}
\end{equation}
the solution of which is explicit for $q=0$ and $q=1$: 
\begin{equation}
	D_0=2, \quad D_1 =1+ \frac{2 \log 2}{ \log (a^{-1} +(1-a)^{-1})}.
\end{equation}
Note that the solution for $q=1$ is obtained with Bernouilli's rule by differentiating \eqref{eq:Fractalbakermap} with respect to $q$, and taking the limit $q\to1$.

\section{Single and multiple bundles statistics.}\label{sec:sampling}
 The variability of a set of random numbers is always greater than the average variability of a subset of these numbers. Thus, the stretching variance \textit{among} bundles of similar sizes-- denoted  $\sigma^2_{\rho^{-1},n}$-- is always larger than the average stretching variance inside the bundle (Eq.~\eqref{eq:variance}). We have
 \begin{equation}
 \sigma^2_{\rho^{-1},n} =\left(\frac{n}{n-1}\right)  \sigma^2_{\rho^{-1}|n}.
 \label{eq:statswhole}
 \end{equation}
For instance, for bundles made of two lamellae, we expect $\sigma^2_{\rho^{-1},n}$ to be twice larger as $\sigma^2_{\rho^{-1}|n}$ given by Eq.~\eqref{eq:variance}. The difference between the statistics of the set and its subset tends to reduce at large $n$, where $\sigma^2_{\rho^{-1}|n}\to \sigma^2_{\rho^{-1},n}$. For $n=1$, both $\sigma^2_{\rho^{-1}|n}$ and $n-1$ cancel out, so that the previous equation is undetermined.

Assuming that bundles of similar size have independent stretching histories, the variability of the aggregated scalar concentration is obtained from independent realisations of the random sum (Eq.~\eqref{eq:aggconcentration}). Thus, the variance of $c|n$ reads
\begin{equation}
\sigma^2_{c|n}=n  \left(\frac{\sqrt{\pi} \theta_0 s_0}{s_a}\right)^2  \sigma^2_{\rho^{-1},n} = \left(\frac{\sqrt{\pi}  \theta_0 s_0}{s_a}\right)^2 \sigma^2_{\rho^{-1}|n}\left(\frac{n^2}{n-1}\right) =\frac{(\sqrt{\pi}  \theta_0 \ell_0 s_0)^2}{\mathcal{A}^2} \left(\frac{n^{2-\tilde{\gamma}} -  1}{n-1} \right).
\label{eq:xi2}
\end{equation}
When $n$ is large, however, we recover 
\begin{equation}
\sigma^2_{c|n} \sim n^{-\tilde{\gamma}+1},
\label{eq:independent2}
\end{equation}

\end{document}